%% file: main.tex
\documentclass[a4paper,12pt,reqno]{amsart}

\usepackage{a4}
\usepackage{verbatim}

\usepackage{color}
\usepackage{comment}
\usepackage{amssymb}
\usepackage{color}
\usepackage{graphicx}
\usepackage{floatflt}
\usepackage{float} 		

\usepackage{amsthm}
\usepackage{amscd}
\usepackage{hyperref}

\newtheorem{definition}{Definition}
\newtheorem{example}[definition]{Example}

\newtheorem{theorem}[definition]{Theorem}
\newtheorem*{theorem*}{Theorem}
\newtheorem{lemma}[definition]{Lemma}
\newtheorem*{lemma*}{Lemma}
\newtheorem{remark}[definition]{Remark}

\newtheorem{corollary}[definition]{Corollary}
\newtheorem*{corollary*}{Corollary}

\newtheorem{proposition}[definition]{Proposition}

\def\XXint#1#2#3{{\setbox0=\hbox{$#1{#2#3}{\int}$}
		\vcenter{\hbox{$#2#3$}}\kern-.5\wd0}}

\makeatletter
\@addtoreset{definition}{section}
\@addtoreset{equation}{section}
\makeatother

\newcommand{\cS}{\mathcal{S}}
\DeclareMathOperator{\Res}{Res}

\allowdisplaybreaks[4]

\title[Symplectic (Non-)Invariance of the Free Energy]{Symplectic (Non-)Invariance of the Free Energy in Topological Recursion}
\author{Alexander Hock}

\address{Institute for Mathematics, University of Heidelberg, Mathematikon, Im Neuenheimer Feld 205,
	69120 Heidelberg, Germany 
{\itshape email address:} \normalfont  
\texttt{alexander.hock@uni-heidelberg.de}}

\begin{document}
\maketitle

\begin{abstract}
Let $F_g$ be the free energy derived from Topological Recursion for a given spectral curve on a compact Riemann surface, and let $F_g^\vee$ be its $x$-$y$ dual, that is, the free energy derived from the same spectral curve with the roles of $x$ and $y$ interchanged. $F_g$ is sometimes called a symplectic invariant due to its invariance under certain symplectomorphisms of the formal symplectic form $dx\wedge dy$. However, the free energy is not generally invariant under the swap of $x$ and $y$; thus, the difference $F_g - F_g^\vee$ is nonzero.

We derive a new formula for this difference for all $g\geq 2$ in terms of a residue calculation at the singularities of $x$ and $y$, including cases where $x$ and $y$ have logarithmic singularities. For the derivation, we apply recent developments from $x$-$y$ duality within the theory of (Logarithmic) Topological Recursion. The derived formulas are particularly useful for spectral curves with a trivial $x$-$y$ dual side, meaning those with vanishing $F_g^\vee$. In such cases, one obtains an explicit result for $F_{g\geq 2}$ itself.

We apply this to several classes of spectral curves and prove, for instance, a recent conjecture by Borot et al. that the free energies $F_g$ computed by Topological Recursion for the "Gaiotto curve" coincide with the perturbative part (in the $\Omega$-background) of the Nekrasov partition function of $\mathcal{N}=2$ pure supersymmetric gauge theory. Similar computations also provide $F_g$ for the CDO curve related to Hurwitz numbers, or the negative $r$-spin curve related to $\Theta$-class intersection numbers on $\overline{\mathcal{M}}_{g,n}$.

\end{abstract}

\section{Introduction and Main Result}
Topological Recursion (TR) is a universal procedure that recursively defines an infinite family of multidifferentials on a Riemann surface \cite{Eynard:2007kz}. Originally, TR was motivated by random matrix theories, but was later more abstractly defined for a given set of initial data, the spectral curve. An important feature of TR is its relation to an incredibly large number of distinguished areas in mathematics, mathematical physics, and theoretical physics. It is almost impossible to list all applications and interconnections of TR. The most prominent ones include enumerative geometry, Hurwitz theory, JT gravity, topological string theory, knot theory, and integrable systems \cite{Eynard:2007kz,Bouchard:2007ys,Eynard:2007fi,Bouchard:2007hi,Bychkov:2020yzy,Saad:2019lba,Borot:2012cw,Alexandrov:2024qfe}. 

The initial data of TR, the spectral curve, is denoted by $(\Sigma,x,y,B)$, where $\Sigma$ is a Riemann surface, $x,y$ are two ramified coverings from $\Sigma$ over $\mathbb{P}^1$, and $B$ is a specific bidifferential on $\Sigma^2$. From this data, the universal definition of TR constructs infinitely many multidifferentials $\omega_{g,n}$ on $\Sigma^n$, indexed by $g\in \mathbb{Z}_{\geq 0}$ and $n\in \mathbb{Z}_{\geq 1}$. Depending on the application, these differentials play a fundamental role. TR itself is recursively defined in $2g+n-2$, meaning that to compute $\omega_{g,n}$, all $\omega_{g',n'}$ with $2g'+n'-2<2g+n-2$ must first be computed. However, examples are known where explicit formulas for $\omega_{g,n}$ (or rather for its application in enumerative geometry for instance) at least for small $n$ can be obtained by other techniques such as Virasoro constraints or orthogonal polynomials. 

Thus, TR is not commonly applied for explicit computations due to its recursive nature. This limitation has historically discouraged mathematicians and physicists from using TR in favor of other technical tools. Nevertheless, TR has still provided deep interconnections between previously distinct areas of mathematics and physics, reducing them to the same universal recursive procedure. However, practical computations with TR have remained challenging.\\

There is a natural duality in the theory of TR by interchanging the roles of the two coverings $x$ and $y$, which gives rise to a second (dual) family $\omega_{g,n}^\vee$ that is by definition different from $\omega_{g,n}$. In some situations where explicit expressions are known in the literature from techniques other than TR, the dual differentials $\omega_{g,n}^\vee$ are trivial, meaning $\omega_{g,n}^\vee = 0$ for $2g+n-2>0$. In the last few years, completely new insights within the theory of TR have been obtained through the derivation of an algebraic formula relating the two families $\omega_{g,n}$ and $\omega_{g,n}^\vee$ \cite{Alexandrov:2023tgl}, see also \cite{Hock:2022wer,Hock:2022pbw}. In the theory of free probability, this duality has an interpretation as a universal functional relation between higher-order free cumulants and moments \cite{Borot:2021thu}. 

With the help of the $x$-$y$ duality relating the two dual families $\omega_{g,n}$ and $\omega_{g,n}^\vee$, it has been possible to recover explicit results directly from TR, for instance, for intersection numbers on the moduli space of complex curves, Gromov-Witten invariants, and Hurwitz numbers \cite{Bychkov:2020yzy,Hock:2023qii}. \\

From the multidifferentials $\omega_{g,n}$, the so-called free energy $F_g$ is obtained. It can also be understood as $\omega_{g,0} = F_g$, defined through a dilaton equation for $g\geq 2$
\begin{align}\label{freeenergy}
    F_g=\frac{1}{2-2g}\sum_{p_i}\Res_{q\to p_i}\Phi(q)\omega_{g,1}(q),
\end{align}
where the $p_i$ are the ramification points of the covering $x$, and $\Phi(q)=\int^q \omega_{0,1}(\bullet)$ is some primitive of $\omega_{0,1}$ (the integration constant being irrelevant for the computation of $F_g$). There is also a definition for $F_g$ with $g\in\{0,1\}$ \cite{Eynard:2007kz} which will not be of interest in this article. The free energy defined by \eqref{freeenergy} is sometimes called symplectic invariants, as it remains invariant under some (but not all) symplectic transformations that preserve the formal symplectic form $dx \wedge dy$. However, one specific symplectic transformation is the prescribed $x$-$y$ duality (up to a sign), which swaps $x$ with $y$. Originally, in the theory of TR, the invariance of $F_g$ under the $x$-$y$ duality was conjectured \cite{Eynard:2007nq}, but this was later found to be incorrect, as observed in \cite{Bouchard:2011ya}. The difference between $F_g$ and its $x$-$y$ dual $F_g^\vee$ was understood in \cite{Eynard:2013csa} for algebraic spectral curves, where $x$ and $y$ are meromorphic functions. However, this result does not extend to functions $x,y$ that admit logarithmic singularities while $dx$ and $dy$ remain meromorphic. These types of spectral curves have important applications in the Gromov-Witten theory of toric Calabi-Yau threefolds \cite{Bouchard:2007ys} or in knot theory \cite{Borot:2012cw}.  

Building on the $x$-$y$ duality, an extension of TR was proposed in \cite{Hock:2023dno} and further developed in \cite{Alexandrov:2023tgl}, now known as Logarithmic Topological Recursion (Log-TR). This extended framework allows for the inclusion of logarithmic behavior in $x$ and $y$ in a way that is consistent with the $x$-$y$ duality. Log-TR reduces to ordinary TR in the absence of logarithmic singularities, or if $dx$ is singular at the logarithmic singularities of $y$. Exactly this situation occurs in the common computation of TR in Gromov-Witten theory, where a generic \textit{framing} is chosen, which ensures that $dx$ is singular at the logarithmic singularities of $y$. The new insight provided by Log-TR actually explains why a framing is necessary, since otherwise the naive application of TR would have yielded incorrect results.

Let $\omega_{g,n}$ be generated by Log-TR (see \eqref{eq:TR-introLog} for the definition) and let the free energy $F_{g\geq 2}$ be defined by \eqref{freeenergy}. Similarly, denote by $\omega_{g,n}^\vee$ the differentials generated by the $x$-$y$ dual spectral curve and by $F_{g\geq 2}^\vee$ the corresponding free energy. The main theorem of the article is the following:
\begin{theorem}\label{thm:maintheorem}
    For a spectral curve with compact Riemann surface, the difference of the free energy $F_g$ and its $x$-$y$ dual $F_g^\vee$ is, for any $g\geq 2$,
    \begin{align}\label{eq:thmformula}
        &(2-2g)(F_g-F_g^\vee)\\\nonumber
        =&\sum_{a_i,a_i^\vee,b}\mathop{\mathrm{Res}}_{z\to a_i,a_i^\vee,b}[\hbar^{2g-1}]dy\sum_{m\geq 2}\frac{d^{m-1}y}{dx^{m-1}}\\\nonumber
        &\times [u^m]\frac{\exp\left(\sum_{h,n}\frac{\hbar^{2h+n-2}}{n!}\int_{y-\hbar u/2}^{y+\hbar u/2}\bigg(\omega_{h,n}^\vee-\frac{\delta_{(h,n),(0,2)}dy_1dy_2}{(y_1-y_2)^2}\bigg)-x u\right)}{u\hbar},
    \end{align}
    where the sum is taken over all $a_i,a_i^\vee$, which are the logarithmic poles of $y$ and $x$ in some local coordinate $z$, respectively, and the points $\{b\}$, which are the singularities of $x,y$ that are not logarithmic. The operation $[z^m]$ extracts the $m$th coefficient of the formal expansion to its right; that is, for $f=\sum_{n}a_nz^n$, one has $[z^n]f=a_n$. The integral expression $\int_{y-\hbar u/2}^{y+\hbar u/2}\omega_{g,n}^\vee$ means that each argument of $\omega_{g,n}^\vee$ is integrated separately from $y-\hbar u/2$ to $y+\hbar u/2$.
\end{theorem}

The RHS of the formula \eqref{eq:thmformula} includes only terms of $\omega_{g',n'}^\vee$ with $2g'+n'-2<2g-2$ due to the $[\hbar^{2g-1}]$ coefficient. In the special situation of a genus-zero spectral curve with unramified $y$, we have $F_g^\vee=0$ since there are no ramification points ($\{p_i^\vee\}=\emptyset$). From the theorem, the following corollary can be concluded, giving an explicit formula for the free energy $F_g$.
\begin{corollary}\label{cor:explicitFg}
    Let $\cS(t)$ be a formal expansion of $\cS(t)=\frac{e^{t/2}-e^{-t/2}}{t}$.
    If $y$ is unramified, one can choose a global coordinate such that $y=z$ or $y=\log z$. If $x$ has logarithmic singularities, they are denoted by $a_i^\vee$ and the residue of $dx$ at $a_i^\vee$ is $\frac{1}{\alpha_i^\vee}$. The free energy $F_{g}$ admits, for $g\geq 2$, the following explicit formula:\\
    $\bullet$ if $y=z$ 
    \begin{align}
        F_g=&\frac{1}{2-2g}\sum_{a_i^\vee,b}\mathop{\mathrm{Res}}_{z\to a_i^\vee, b}[\hbar^{2g-1}]dy\sum_{m\geq 2}\frac{d^{m-1}y}{dx^{m-1}}\\\nonumber
        &\times [u^m]\frac{\exp\left[\big(\cS(u \hbar \partial_z)-1\big)x u+u\cS(u \hbar \partial_z)\sum_i\big(\frac{1}{\cS(\alpha_i^\vee\hbar \partial_z)}-1\big)\frac{\log(z-a_i^\vee)}{\alpha_i^\vee}\right]}{u\hbar}
    \end{align}
    $\bullet$  if $y=\log z$  
    \begin{align}\label{coreqylog}
        F_g=&\frac{1}{2-2g}\sum_{a_i^\vee,b}\mathop{\mathrm{Res}}_{z\to 0,\infty,a_i^\vee,b}[\hbar^{2g-1}]dy\sum_{m\geq 2}\frac{d^{m-1}y}{dx^{m-1}}\\\nonumber
        &\times [u^m]\frac{\exp\left[\big(\cS(u \hbar \partial_{y(z)})-1\big)x u+u\cS(u \hbar \partial_{y(z)})\sum_i\big(\frac{1}{\cS(\alpha_i^\vee\hbar \partial_{y(z)})}-1\big)\frac{\log(z-a_i^\vee)}{\alpha_i^\vee}\right]}{e^{\hbar u /2}-e^{-\hbar u /2}}.
    \end{align}
\end{corollary}

Applying the formula of Corollary \ref{cor:explicitFg} to the spectral curve studied in \cite{Borot:2024uos} after a change of variables, we are able to obtain all $F_{g\geq 2}$ and thus prove a conjecture by Borot et al. \cite{Borot:2024uos} that the free energy of TR for the corresponding curve coincides with the perturbative part of the Nekrasov partition function of $\mathcal{N}=2$ pure supersymmetric gauge theory \cite{Nekrasov:2003rj}. This demonstrates the power of the $x$-$y$ duality in obtaining explicit results in TR. 

We will apply the same formalism to the spectral curve studied in \cite{chidambaram2024bhurwitznumberswhittakervectors} (with $b=0$ in their notation) in relation to hypergeometric Hurwitz numbers. 

However, applying the formula from Corollary \ref{cor:explicitFg} can be quite tricky. Numerical tests have verified several additional examples, such as the topological vertex curve \cite{Aganagic:2003db}, the resolved conifold, and generalizations of the conifold related to the asymptotic expansion of the MacMahon function. With our new method, the free energy can be computed explicitly, and this will be studied elsewhere.

The results of Theorem \ref{thm:maintheorem} and Corollary \ref{cor:explicitFg} also hold in a more general setting. By the deformation theory and limiting procedures studied in \cite{Alexandrov:2024tjo}, the formulas of Theorem \ref{thm:maintheorem} and Corollary \ref{cor:explicitFg} hold for higher order TR where $y$ is required to be regular at the ramification points of $x$ and vice versa. In this situation, the definition of higher order TR by Bouchard and Eynard \cite{Bouchard:2012yg} coincides with the definition of generalized TR from \cite{Alexandrov:2024tjo}, which is essentially constructed from the $x$-$y$ duality and thus applicable here.

\section*{Acknowledgments}
I am grateful to Ondra Hulik, Raphael Senghaas, Valdo Tatitscheff, and Johannes Walcher for discussions related to physics applications, and also Nitin Kumar Chidambaram, Jakob Lindner and Olivier Marchal for further comments and suggestions. This work was supported through the project ``Topological Recursion, Duality and Applications''\footnote{``Funded by the Deutsche Forschungsgemeinschaft (DFG, German Research Foundation) -- Project-ID 551478549''}.

\section{Logarithmic Topological Recursion and $x$-$y$ Duality}
This section provides background on recent developments in the theory of TR, presenting new tools for practical computation. We will start directly with the definition of the so-called Logarithmic Topological Recursion (Log-TR). This is an extension of the original TR by properly including logarithmic singularities. A fundamental property of Log-TR, not satisfied by TR, is its compatibility with the $x$-$y$ duality transformation formulas if logarithmic singularities are present. The $x$-$y$ duality formulas will be used for the derivation of the main theorems.

\subsection{Logarithmic Topological Recursion}
Let us first recall the set-up in Log-TR \cite{Alexandrov:2023tgl}. The input is the spectral curve, a tuple $(\Sigma, x, y, B)$, where $\Sigma$ is a \textup{compact}\footnote{For the general definition of TR and Log-TR, compactness does not have to be assumed, but it will be important for us in this article.} Riemann surface, and $x, y: \Sigma \to \mathbb{C}$ are functions on $\Sigma$ with simple\footnote{Simple in the sense that two branches come together at this point.} ramification points, and $dx, dy$ are meromorphic, allowing $x$ and $y$ to have local logarithmic singularities\footnote{The logarithmic singularity is not denoted as a ramification point. We distinguish between logarithmic and algebraic ramification points.}. The bidifferential $B$ is the Bergman kernel; that is, $B$ is symmetric, has a double pole on the diagonal and no other poles, and is normalized along the $\mathcal{A}$-cycles of $\Sigma$ for a given choice of homology cycles. The set $Ram(x)$ is the set of ramification points of $x$. The points in $Ram(x)$ are denoted by $p_i$. At a (simple) ramification point $p_i \in Ram(x)$, there is a unique Deck transformation defined by $\sigma_i$ with $x(q)=x(\sigma_i(q))$ and with $p_i$ as the fixed point, i.e. $\sigma_i(p_i)=p_i$. We denote the logarithmic poles of $y$ (which are not simultaneously poles of $dx$) by $a_1,\ldots,a_M$ and refer to them as log-vital singular points. The logarithmic poles of $y$ which are poles of $dx$ can be neglected. At the points $a_1,\ldots,a_M$, the differential $dy$ has nonzero residues $\frac{1}{\alpha_1},\ldots,\frac{1}{\alpha_M}$, respectively. 

Then, the multidifferentials of Log-TR are defined by $\omega_{0,1}=y\,dx$, $\omega_{0,2}=B$, and, for negative Euler characteristic $\chi=2-2g-n<0$, recursively by
\begin{align}
  \label{eq:TR-introLog}
&  \omega_{g,n+1}(I,z)
  :=\sum_{p_i\in Ram(x)}
  \Res\displaylimits_{q\to p_i}
  K_i(z,q)\bigg(
   \omega_{g-1,n+2}(I, q,\sigma_i(q))\\
  &\qquad \qquad\qquad\qquad\qquad\qquad
  +
   \sum_{\substack{g_1+g_2=g\\ I_1\sqcup I_2=I\\
            (g_i,I_i)\neq (0,\emptyset)}}
   \omega_{g_1,|I_1|+1}(I_1,q)
  \omega_{g_2,|I_2|+1}(I_2,\sigma_i(q))\!\bigg)\nonumber\\\nonumber
  &\qquad \qquad+\delta_{n,0}\sum_{i=1}^M\Res\displaylimits_{q\to a_i}\int_{a_i}^q \omega_{0,2}(z, \bullet)[\hbar^{2g}]\left(\frac{1}{\alpha_i S(\alpha_i \hbar \partial_{x(q)})}\log (q-a_i)\right)dx(q).
\end{align}
Here, the recursion kernel is 
\[
K_i(z,q)=\frac{\frac{1}{2}\int^{q}_{\sigma_i(q)}\omega_{0,2}(z,\bullet)}{\omega_{0,1}(q)-\omega_{0,1}(\sigma_i(q))}.
\]
With $I=\{z_1,\ldots,z_n\}$, we denote the set of coordinates on the different copies of $\Sigma^n$.

The difference between Log-TR and the original definition of TR by Eynard and Orantin in \cite{Eynard:2007kz} is the last line in \eqref{eq:TR-introLog}. Note that this term occurs only if $n=0$, that is, it contributes only to $\omega_{g,1}$ (but recursively to all $\omega_{g,n}$), and only if $y$ has a logarithmic pole $a_i$ that is not a pole of $dx$. Due to the projector $\Res_{q\to p_i} \int\omega_{0,2}(z,\bullet)$, all multidifferentials $\omega_{g,n}$ for $2g+n-2>0$ have poles only at the ramification points $p_i\in Ram(x)$, and $\omega_{g,1}$ additionally has poles at the log-vital singular points of $y$, namely $a_1,\ldots,a_M$. The scaling property is satisfied, that is, rescaling $\omega_{0,1}\to \lambda \omega_{0,1}$ rescales all $\omega_{g,n}\to \lambda^{2-2g-n}\omega_{g,n}$. Furthermore, the $\omega_{g,n}$ of Log-TR satisfy the linear and quadratic loop equations \cite{Alexandrov:2023tgl}.

\begin{remark}\label{rmk:logvitalpoints}
    The log-vital singular points are logarithmic singularities of $y$ which are \textit{not} poles of $dx$. Including the logarithmic poles of $y$ that are poles of $dx$ in the definition of Log-TR does not change the definition, since at those points the residue calculation in the last line of \eqref{eq:TR-introLog} vanishes and, thus, does not contribute.
\end{remark}

\begin{remark}
One important application of the original formulation of TR was on the $B$-model side in topological string theory for toric Calabi-Yau threefolds, known as the remodeling the $B$-model conjecture \cite{Bouchard:2007ys}. The spectral curve is the mirror curve of a toric Calabi-Yau threefold, and the corresponding functions $x,y$ have logarithmic singularities. For the remodeling conjecture to hold, a so-called generic framing $f$ needs to be assumed. From the TR perspective, the framing is a deformation of the curve by adding multiples of $y$ to $x$, that is, for TR to work one needs the transformation of $(x,y)\mapsto (\tilde{x},\tilde{y})$ by
\begin{align}\label{framing}
    &x\mapsto \tilde{x}=x+ f y,\qquad f\in\mathbb{Z},\\\nonumber
    &y\mapsto \tilde{y}=y.
\end{align}
If $y$ has log-vital points $a_i$, which are not poles of $dx$, then the framing transformation \eqref{framing} yields a curve with no log-vital points of $y$ anymore. To be more precise, if $y$ is locally at $a_i$ of the form $\frac{1}{\alpha_i}\log(z-a_i)$ and $x$ is regular at $a_i$, the transformed $\tilde{x}=x+fy$ is locally of the form $\frac{f}{\alpha_i}\log(z-a_i)+\text{regular}$. Therefore, $d\tilde{x}$ now indeed has a pole at $a_i$ (and actually, for generic $f$, at all $a_i$). In this situation, as mentioned in Remark \ref{rmk:logvitalpoints}, Log-TR reduces to the original TR. However, it was important for the remodeling conjecture to have a framing; otherwise, TR does not compute the correct correlators of the $B$-model. \textbf{Log-TR explains why the framing was inevitable for the remodeling conjecture from the perspective of TR, and how to extend TR (to Log-TR) to include all (also non-generic) framings.}
\end{remark}

\subsection{$x$-$y$ duality}
Log-TR arose from the idea to extend the $x$-$y$ duality to spectral curves including logarithmic singularities. It was observed in \cite{Hock:2023dno} that, when taking the original formulation of TR, the $x$-$y$ duality was not satisfied. What does this statement actually mean? 

In the theory of TR, we start with the spectral curve $(\Sigma,x,y,B)$ and generate the infinite family of multidifferentials $\omega_{g,n}$. Changing $x$ and $y$, i.e. considering the spectral curve $(\Sigma,y,x,B)$, generates another infinite family of multidifferentials $\omega_{g,n}^\vee$ which are different from $\omega_{g,n}$. A long-standing question in the theory of TR was about the relation between $\omega_{g,n}$ and $\omega_{g,n}^\vee$. A conjecture about this relation was stated in \cite{Borot:2021thu}, proved for $g=0$ under the assumption of a loop insertion operator in \cite{Hock:2022wer}, and finally settled for compact curves and meromorphic $x,y$ in \cite{Alexandrov:2022ydc}; see also \cite{JEP_2022__9__1121_0,Hock:2022pbw} for further references. The statement can be summarized as
\begin{align}\label{dualxyexpression}
    \omega_{g,n} = \text{Expr}_{g,n}\bigg(\big\{\omega_{h,m}^\vee\big\}_{2h+m-2 \leq 2g+n-2}, \{dy_i, dx_i\}_{i=1,\dots,n}, \big\{\frac{dy_i\,dy_j}{(y_i-y_j)^2}\big\}_{i,j=1,\dots,n}\bigg)
\end{align}
where $\text{Expr}_{g,n}$ is an algebraic combinatorial expression depending on $\omega_{h,m}^\vee$, the differentials $dx_i,dy_i$, and $\frac{dy_i\,dy_j}{(y_i-y_j)^2}$, which serves to regularize $\omega_{0,2}^\vee$ at the diagonal.  

The actual expression $\text{Expr}_{g,n}$ relating the two families has a rather involved combinatorial nature and becomes increasingly complicated for higher $n$. It consists of a differential operator acting on $\omega_{h,m}^\vee$, $dx_i$, and $dy_i$. We will not cite the most general relation between $\omega_{g,n}$ and $\omega_{g,n}^\vee$ but rather refer to the articles $op.\ cit.$ 

Testing the universal relation (or expression) between the two families $\omega_{g,n}$ and $\omega_{g,n}^\vee$, in the case where $x$ and $y$ are allowed to have logarithmic singularities, turns out to fail. However, enforcing the universal relation to hold motivated the definition of Log-TR. Thus, we can reformulate the result of \cite{Alexandrov:2023tgl} as
\begin{theorem}[\cite{Alexandrov:2023tgl}]
    Let $\omega_{g,n}$ and $\omega_{g,n}^\vee$ be generated by Log-TR \eqref{eq:TR-introLog}; then both families are related by the universal $x$-$y$ duality \eqref{dualxyexpression}.
\end{theorem}

Strikingly, the expression of the duality depends not explicitly on $x$ or $y$ but rather on $dx$ and $dy$, which, even for logarithmic singularities of $x,y$, remain meromorphic forms. The dependence on $\frac{dy_i\,dy_j}{(y_i-y_j)^2}$ is rather implicit, since it is used to regularize $\omega_{0,2}^\vee$. Since it is a duality, the same expression \eqref{dualxyexpression} holds when interchanging $\omega_{g,n}$ with $\omega_{g,n}^\vee$ and $x$ with $y$. 

We do not wish to make a great fuss about the universality and importance of the $x$-$y$ duality—which has already been applied to prove an open conjecture concerning integrability \cite{Alexandrov:2024hgu,Alexandrov:2024qfe} and will have further significant impact—but rather we aim to apply it for actual computations of the free energy. To that end, we will need the simplest version of \eqref{dualxyexpression} in the special case $(g,n)=(g,1)$, namely $\text{Expr}_{g,1}$.

Let $\cS(t)=\frac{e^{t/2}-e^{-t/2}}{t}$. We write $[t^m]$ as an operator extracting the $m$th coefficient of the formal expansion to its right, that is, for $f=\sum_{n}a_nt^n$ this is $[t^n]f=a_n$. We denote $d\frac{1}{dx}$ as an operation on a 1-form $\omega$ producing a 1-form, by first contracting with respect to $dx$ and then taking the exterior derivative. The expression of $\omega_{g,1}$ in terms of $\omega_{h,m}^\vee$ with $2h+m-2\leq 2g-1$ reads (see for instance \cite{Alexandrov:2022ydc,Hock:2022pbw,Alexandrov:2023tgl,Hock:2023qii})
\begin{align}\label{xyomg1}
    &\omega_{g,1}=-[\hbar^{2g-1}]\sum_{m\geq 0}\left(-d\frac{1}{dx}\right)^m [u^m]dy\\\nonumber
    &\times\frac{\exp\left(\sum_{h\geq 0,m\geq 1}\frac{\hbar^{2h+m-2}}{m!}\bigg(\bigg{\vert}_{\tilde{y}_i\to y}\!\!\!\!\!\!\!\!\!\!\prod_{i=1}^m \hbar u \cS(u\hbar \frac{d}{d\tilde{y}_i})\frac{1}{d\tilde{y}_i}\bigg)\big(\omega_{h,m}^\vee-\frac{\delta_{(h,m),(0,2)}d\tilde{y}_1d\tilde{y}_2}{(\tilde{y}_1-\tilde{y}_2)^2}\big)-x u\right)}{u\hbar}.
\end{align}
Let us comment on a few properties of this very special form of the $x$-$y$ duality formula. Due to the $[\hbar^{2g-1}]$ coefficient, only finitely many terms in the sum over $m$ contribute. Each summand in the sum over $m$ extracts the $u^m$ coefficient, which acts by the $m$th power of the operator $-d\frac{1}{dx}$. The multidifferential $\omega_{h,m}^\vee=\omega_{h,m}^\vee(\tilde{y}_1,\dots,\tilde{y}_m)$ in the exponential is contracted with respect to $d\tilde{y}_i$ in each variable with an additional action of $\cS(u\hbar \frac{d}{d\tilde{y}_i})$ on top. For $(h,m)=(0,1)$, $\omega_{0,1}^\vee$ is the leading term in $\hbar$. This leading term $\frac{1}{\hbar}\hbar u \cS(u\hbar \frac{d}{dy})x$ is annihilated by $-xu$ at leading order in $\hbar$, thus there is no $\hbar$ with negative or constant power inside the exponential. For $(h,m)=(0,2)$, $\omega_{0,2}^\vee$ is regularized by $\frac{\delta_{(h,m),(0,2)}d\tilde{y}_1d\tilde{y}_2}{(\tilde{y}_1-\tilde{y}_2)^2}$ to have a well-defined diagonal by the limit $\big{\vert}_{\tilde{y}_i\to y}$ for $\tilde{y}_1=\tilde{y}_2=y$.

At a fixed order $[\hbar^{2g-1}]$, the leading order in the $u$-expansion is given by 
\begin{align*}
    &[u^0]:&&-\omega_{g,1}^\vee,\\
    &[u^1]:&& -\frac{1}{dy}\bigg(\omega^\vee_{g-1,2}+\frac{1}{2}\sum_{\substack{g_1+g_2=g\\ g_i>0}}\omega^\vee_{g_1,1}\omega^\vee_{g_2,1}\bigg).
\end{align*}
For $[u^1]$, this expression is still a 1-form, since the expression in the parentheses is a bidifferential at the diagonal. For $g=1$, $\omega_{0,2}$ would have been regularized. Inserting this back into \eqref{xyomg1}, the $[u^1]$-term gets an additional action of $-d\frac{1}{dx}$. We can write the first orders as
\begin{align*}
    \omega_{g,1} &= -\omega_{g,1}^\vee + d\bigg(\frac{\omega_{g-1,2}^\vee+\frac{1}{2}\sum_{\substack{g_1+g_2=g\\ g_i>0}}\omega_{g_1,1}^\vee\,\omega_{g_2,1}^\vee}{dx\, dy}\bigg) + \mathcal{O}\bigg(\big(d\frac{1}{dx}\big)^2\Omega\bigg),
\end{align*}
where $\mathcal{O}\bigg(\big(d\frac{1}{dx}\big)^2\Omega\bigg)$ denotes the remaining terms for all $m\geq 2$ with at least a second-order operation of $d\frac{1}{dx}$ on some 1-form $\Omega$. 

Let us also comment on the function $\cS(t)$, which acts on the multidifferential via a differential operator. The leading orders are $\cS(t)=1+\frac{t^2}{24}+\mathcal{O}(t^4)$. However, $\cS(u\hbar \frac{d}{dy})\frac{1}{dy}$ can also be interpreted as a Riemann-Hilbert-like problem in the following formal sense. Let $\omega$ be a 1-form; then we can write formally
\begin{align*}
    u\hbar\cS(u\hbar \frac{d}{dy})\frac{1}{dy}\omega 
    &= \frac{e^{\frac{u\hbar}{2} \frac{d}{dy}}-e^{-\frac{u\hbar}{2} \frac{d}{dy}}}{\frac{d}{dy}}\frac{1}{dy}\omega\\[1mm]
    &= \Big(e^{\frac{u\hbar}{2} \frac{d}{dy}}-e^{-\frac{u\hbar}{2} \frac{d}{dy}}\Big)d^{-1}\omega\\[1mm]
    &= \Big(e^{\frac{u\hbar}{2} \frac{d}{dy}}-e^{-\frac{u\hbar}{2} \frac{d}{dy}}\Big)\int^y \omega 
    = \int_{y-\frac{u\hbar}{2}}^{y+\frac{u\hbar}{2}}\omega,
\end{align*}
as a formal power series in $\hbar$. If $\Phi(y)=\int^y \omega $ is a primitive of $\omega$, then the expression can be understood equivalently as
\begin{align*}
    u\hbar\cS(u\hbar \frac{d}{dy})\frac{1}{dy}\omega = \Phi\Big(y+\frac{\hbar u}{2}\Big)-\Phi\Big(y-\frac{\hbar u}{2}\Big),
\end{align*}
which is indeed of the form of a Riemann-Hilbert problem at $\hbar=0$.

Inserting $u\hbar\cS(u\hbar \frac{d}{dy})\frac{1}{dy}$ in terms of an integral representation, the duality formula for $\omega_{g,1}$ can be written compactly as
\begin{align}\label{xyomg1int}
    \omega_{g,1}&=[\hbar^{2g-1}]\sum_{m\geq 0}\left(-d\frac{1}{dx}\right)^m [u^m](-dy)\\\nonumber
    &\times\frac{\exp\left(\sum_{h\geq 0,m\geq 1}\frac{\hbar^{2h+m-2}}{m!}\bigg(\int_{y-\frac{u\hbar}{2}}^{y+\frac{u\hbar}{2}} \big(\omega_{h,m}^\vee-\frac{\delta_{(h,m),(0,2)}d\tilde{y}_1d\tilde{y}_2}{(\tilde{y}_1-\tilde{y}_2)^2}\big)-x u\right)}{u\hbar}.
\end{align}
Here, we understand $\int_{y-\frac{u\hbar}{2}}^{y+\frac{u\hbar}{2}} \omega_{h,m}^\vee$ as an integral on each copy of the Riemann surface $\Sigma$ of $\Sigma^m$, since $\omega_{h,m}^\vee$ lives on $\Sigma^m$ where the projection to each copy is a 1-form.

Let us mention once more that \eqref{xyomg1int} holds in the context of Log-TR \eqref{eq:TR-introLog} if log-vital singular points exist (for $x$ or $y$), and for TR (in the original formulation \cite{Eynard:2007kz}) if no log-vital points exist.\\

Let us discuss some particular examples in which the family $\omega_{g,n}^\vee$ has a trivial form. This is the case if $y$ is unramified. For a large class of genus zero spectral curves, $y$ can be chosen to have the form $y=z$ or $y=\log z$ locally (and thus globally). In both cases, the explicit form of $\omega_{g,n}^\vee$ additionally depends on whether or not $x$ has log-vital points. 
\begin{example}\label{ex:omg1yzxrational}
    Let $y=z$ and assume that $x$ has no log-vital points. Since $y$ has no ramification points ($\{p_i^\vee\}=\emptyset$), the first two lines of the $x$-$y$ dual of \eqref{eq:TR-introLog} do not contribute. Since $x$ has no log-vital points, the third line of \eqref{eq:TR-introLog} also does not contribute. Therefore, all $\omega_{g,n}^\vee=0$ for $2g+n-2>0$ and $\omega_{0,2}^\vee-\frac{dy_1dy_2}{(y_1-y_2)^2}=0$ since $y=z$. The only non-trivial term in \eqref{xyomg1int} comes from $\omega_{0,1}^\vee$. The explicit expression of $\omega_{g,1}$ becomes
    \begin{align*}
        \omega_{g,1}(z)&=[\hbar^{2g-1}]\sum_{m\geq 0}\left(-d\frac{1}{dx(z)}\right)^m [u^m](-dz)\frac{\exp\left(\frac{1}{\hbar}\int_{z-\frac{u\hbar}{2}}^{z+\frac{u\hbar}{2}} x(z)\,dz-x(z) u\right)}{u\hbar}.
    \end{align*}
\end{example}

\begin{example}\label{ex:omg1yzxlog}
    Let $y=z$ and suppose that $x$ has log-vital points $a_i^\vee$ with residue $\Res_{z\to a_i^\vee}dx=\frac{1}{\alpha^\vee_i}$. Since $y$ has no ramification points ($\{p_i^\vee\}=\emptyset$), the first two lines of the $x$-$y$ dual of \eqref{eq:TR-introLog} do not contribute. Since $x$ has log-vital points, the third line of \eqref{eq:TR-introLog} is non-trivial. All $\omega_{g,1}^\vee$ are explicitly given for $g>0$ by 
    \begin{align*}
        \omega_{g,1}^\vee(z)=[\hbar^{2g}]dz\sum_{i}\frac{1}{\alpha_i^\vee\cS(\alpha_i^\vee\hbar \partial_z)}\log(z-a_i^\vee)
    \end{align*}
    and $\omega_{0,1}^\vee(z)=x(z)dz$. Inserting this into \eqref{xyomg1}, all $\omega_{g,1}$ read
    \begin{align*}
        \omega_{g,1}(z)&=[\hbar^{2g-1}]\sum_{m\geq 0}\left(-d\frac{1}{dx(z)}\right)^m [u^m](-dz)\\
        &\times \frac{\exp\left[\big(\cS(u \hbar \partial_z)-1\big)x(z) u+u\cS(u \hbar \partial_z)\sum_i\big(\frac{1}{\cS(\alpha_i^\vee\hbar \partial_z)}-1\big)\frac{\log(z-a_i^\vee)}{\alpha_i^\vee}\right]}{u\hbar}.
    \end{align*}
    One can replace the action of $u \cS(u \hbar \partial_z)$ by the operator $\frac{1}{\hbar}\int_{z-\frac{\hbar u}{2}}^{z+\frac{\hbar u}{2}} dz$.
\end{example}

\begin{example}\label{ex:omg1ylogxrational}
    Let $y=\log z$ and suppose that $x$ has no log-vital points. In this situation all $\omega_{g,n}^\vee=0$ for $2g+n-2>0$. We have non-trivial $\omega_{0,1}^\vee$ and $\omega_{0,2}^\vee$. The computation of $\omega_{0,2}^\vee$ at the diagonal reads
    \begin{align*}
        &\frac{1}{2}\int_{y-\frac{u\hbar}{2}}^{y+\frac{u\hbar}{2}}\int_{y-\frac{u\hbar}{2}}^{y+\frac{u\hbar}{2}}\bigg(\frac{dz_1dz_2}{(z_1-z_2)^2}-\frac{\frac{dz_1}{z_1}\frac{dz_2}{z_2}}{(\log z_1-\log z_2)^2}\bigg)\\[1mm]
        =&\frac{1}{2}\log\bigg(\frac{e^{y_1}-e^{y_2}}{y_1-y_2}\bigg)\Bigg\vert_{y-\frac{u\hbar}{2}}^{y+\frac{u\hbar}{2}}\\[1mm]
        =&\frac{1}{2}\bigg(\log e^{y+\frac{u\hbar}{2}}+\log e^{y-\frac{u\hbar}{2}}-\log e^{y}\frac{e^{\frac{u\hbar}{2}}-e^{-\frac{u\hbar}{2}}}{u\hbar}-\log e^{y}\frac{e^{\frac{u\hbar}{2}}-e^{-\frac{u\hbar}{2}}}{u\hbar}\bigg)\\[1mm]
        =&-\log \cS(u\hbar).
    \end{align*}
    Inserting this into \eqref{xyomg1int}, all $\omega_{g,1}$ read
    \begin{align*}
        \omega_{g,1}(z)&=[\hbar^{2g-1}]\sum_{m\geq 0}\left(-d\frac{1}{dx(z)}\right)^m [u^m](-dy(z))\frac{\exp\left(\frac{1}{\hbar}\int_{y(z)-\frac{u\hbar}{2}}^{y(z)+\frac{u\hbar}{2}} x(z)dy(z)-x(z) u\right)}{e^{\frac{u\hbar}{2}}-e^{-\frac{u\hbar}{2}}}.
    \end{align*}
\end{example}

\begin{example}\label{ex:omg1ylogxlog}
    Let $y=\log z$ and suppose that $x$ has log-vital points $a_i^\vee$ with residue $\Res_{z\to a_i^\vee}dx=\frac{1}{\alpha^\vee_i}$. Similarly to Example \ref{ex:omg1ylogxrational}, there is a non-trivial contribution from $\omega_{0,2}^\vee$. Analogous to Example \ref{ex:omg1yzxlog}, there is also a contribution to $\omega_{g,1}^\vee$. In summary, all $\omega_{g,1}$ read
    \begin{align*}
        \omega_{g,1}(z)&=[\hbar^{2g-1}]\sum_{m\geq 0}\left(-d\frac{1}{dx(z)}\right)^m [u^m](-dy(z))\\
        &\times \frac{\exp\left[\big(\cS(u \hbar \partial_{y(z)})-1\big)x(z) u+u\cS(u \hbar \partial_{y(z)})\sum_i\bigg(\frac{1}{\cS(\alpha_i^\vee\hbar \partial_{y(z)})}-1\bigg)\frac{\log(z-a_i^\vee)}{\alpha_i^\vee}\right]}{e^{\frac{u\hbar}{2}}-e^{-\frac{u\hbar}{2}}}.
    \end{align*}
\end{example}

\section{Free Energy $F_g$}
In the theory of TR, the free energy (sometimes erroneously called \textit{symplectic invariance}) plays a profound role in the theory of TR. It is for $g\geq 2$ is defined by
\begin{align}\label{freeenergy2}
    F_g=\frac{1}{2-2g}\sum_{p_i\in Ram(x)}
    \Res\displaylimits_{q\to p_i} \Phi(q)\omega_{g,1}(q),
\end{align}
where $\Phi(q)=\int^q \omega_{0,1}(\bullet)$ is some primitive of $\omega_{0,1}$; the integration constant does not matter for the computation of $F_g$ since all $\omega_{g,n}$ are residue free. This definition is the original one from \cite{Eynard:2007kz}, even though we are dealing with the extended version of Log-TR. The reader might wonder why, in Log-TR, the poles at $a_i$ in \eqref{freeenergy2} are completely neglected. Extending the sum over all ramification points naively to include the points $a_i$ does not work, since the residue calculation of $\Phi(q)\omega_{g,1}(q)$ at the points $a_i$ does not make sense: the primitive $\Phi(q)$ has logarithmic singularities there. Thus, we keep the definition of the free energy from the original TR for Log-TR.

In fact, for later applications we will deal with spectral curves where no log-vital singular points occur for $y$, so the definitions of $\omega_{g,n}$ and the free energy $F_g$ in Log-TR reduce to those of TR. Why do we then care about Log-TR? The reason is that the $x$-$y$ dual side may have non-trivial log-vital points; in other words, $x$ might have log-vital points. Since including these log-vital points is important to satisfy the universal $x$-$y$ duality transformations, we had to give the general definition.

For $F_g$, one of the main properties, among others, is its invariance under certain deformations of the spectral curve. These deformations correspond to deformations of the formal symplectic form $dx\wedge dy\to d\tilde{x}\wedge d\tilde{y}$. The $x$-$y$ dual free energy is defined by
\begin{align}\label{freeenergydual}
     F^\vee_g=\frac{1}{2-2g}\sum_{p_i^\vee\in Ram(y)}
     \Res\displaylimits_{q\to p_i^\vee} \Phi^\vee(q)\omega^\vee_{g,1}(q),
\end{align}
with $\Phi^\vee(q)=\int^q \omega^\vee_{0,1}(\bullet)=\int^q x(\bullet)dy(\bullet)$.

It was claimed in \cite{Eynard:2007nq} that $F_g$ is also invariant under the specific symplectic transformation of interchanging the roles of $x$ and $y$, that is, $F_g=F_g^\vee$. It turned out not to hold in general \cite{Bouchard:2011ya,Bouchard:2012an} and depends on the poles of $x$ and $y$. For meromorphic $x$ and $y$, a new invariant definition was given in \cite{Eynard:2013csa}
\begin{align}\label{freeenergytrivial}
    \frac{1}{2-2g}\sum_{p_i\in Ram(x)}
    \Res\displaylimits_{q\to p_i} \Phi(q)\omega_{g,1}(q)+\frac{1}{2-2g}\sum_{b}t_i\int_\star^b  \omega_{g,1},
\end{align}
where $b$ are the poles of $x,y$ and $t_i=\Res_{b}\omega_{0,1}$. This new definition turns out to be invariant under swapping $x$ and $y$ in the case of meromorphic $x,y$. However, it does not extend to allow logarithmic singularities of $x$ and $y$. Furthermore, the definition proposed in \cite{Eynard:2013csa} actually trivializes some parts of the free energy.

For instance, the Gaussian unitary ensemble can be represented by the genus-zero spectral curve with $x=z+\frac{1}{z}$ and $y=z$. The free energy has the explicit form 
\[
F_{g\geq 2}=\frac{B_{2g}}{2g(2g-2)},
\]
which is of high importance for the connection to the Euler characteristic of the moduli space of complex curves $\chi(\mathcal{M}_g)$ by Harer and Zagier \cite{Zagier1986}. It is quite obvious that the $x$-$y$ dual free energy $F_g^\vee=0$ for this curve is trivial (since there are no ramification points for $y$). So what does the proposed definition \eqref{freeenergytrivial} of \cite{Eynard:2013csa} do for the Gaussian unitary ensemble? It is almost obvious that \eqref{freeenergytrivial} gives for the Gaussian unitary ensemble the invariant $F_g=F_g^\vee=0$, thus trivializing $F_g$ rather than correcting $F_g^\vee$ to preserve the important connection to $\chi(\mathcal{M}_g)$.

Thus, there are two main problems with the $x$-$y$ duality of the free energy in the theory of TR:
\begin{itemize}
    \item logarithmic singularities are not included,
    \item the proposal \eqref{freeenergytrivial} trivializes some parts of the free energy.
\end{itemize}

Since at the time the articles \cite{Bouchard:2011ya,Bouchard:2012an,Eynard:2013csa} were written, the $x$-$y$ duality was not developed and Log-TR was not defined yet, so the technique was missing to tackle these problems. We will shed new light on both problems and provide a new qualitative understanding of the free energy in TR in general.

For the analysis, we need to distinguish between the following particular sets of points which will be important in the computation of the free energy:
\begin{align*}
    &p_i &&:\text{ramification points of } x,\text{ i.e. } dx(p_i)=0,\\[1mm]
    &p_i^\vee &&:\text{ramification points of } y,\text{ i.e. } dy(p_i^\vee)=0,\\[1mm]
    &a_i &&:\text{log-vital singular points of } y,\\[1mm]
    &a_i^\vee &&:\text{log-vital singular points of } x,\\[1mm]
    &b_i &&:\text{singular (but not logarithmic) points of } y \text{ such that } \Res_{b_i}y\,dx\neq 0,\\[1mm]
    &b_i^\vee &&:\text{singular (but not logarithmic) points of } x \text{ such that } \Res_{b_i^\vee}x\,dy\neq 0.
\end{align*}

In Theorem \ref{thm:mainthmtext} and Corollary \ref{cor:explicitFg}, the sets $\{b_i\}$ and $\{b_i^\vee\}$ are combined into $\{b\}$.

\subsection{Difference between $F_g$ and $F_g^\vee$}
Since we want to extend the analysis to logarithmic singular points for $x$ and $y$ by using Log-TR, we will need to extend certain properties valid for the $\omega_{g,n}$ generated by Log-TR \eqref{eq:TR-introLog}.
\begin{lemma}\label{lem:xyomegag1}
    Let $\omega_{g,n}$ be generated by Log-TR, then the following holds
    \begin{align}\label{lemmaequationxy}
        \sum_i\Res_{z\to p_i} x(z)y(z)\omega_{g,1}(z)=\frac{1}{2}\sum_{j}\mathrm{Res}_{z\to a_j}\frac{x(z)}{dx(z)}\sum_{g_1+g_2=g,\; g_i>0}\omega_{g_1,1}(z)\omega_{g_2,1}(z),
    \end{align}
    where $\{p_i\}$ is the set of all ramification points of $x$ and $\{a_j\}$ the set of all log-vital points of $y$. In case of absence of log-vital points $\{a_i\}=\emptyset$, the property \eqref{lemmaequationxy} reduces to \cite[eq. (4-29)]{Eynard:2007kz}.
    \begin{proof}
        Before starting the computation, we need to recall that all $\omega_{g,n}$ generated by Log-TR satisfy the linear and quadratic loop equation \cite{Alexandrov:2023tgl} for $2g+n-2>0$. This is a local property around a ramification point $p_i$ of $x$ including the local Deck transformation $\sigma_i$ around the ramification point. Let $I=\{z_1,\dots,z_{n}\}$; then the linear loop equation is
        \begin{align*}
            \omega_{g,n+1}(z,I)+\omega_{g,n+1}(\sigma_i(z),I)\in \mathcal{O}((z-p_i)^1)dz
        \end{align*} 
        and the quadratic loop equation is
        \begin{align*}
            \sum_{\substack{J_1\sqcup J_2=I \\ g_1+g_2=g}}\omega_{g_1,|J_1|+1}(z,J_1)\,\omega_{g_2,|J_2|+1}(\sigma_i(z),J_2)+\omega_{g-1,2}(z,\sigma_i(z),I)\in \mathcal{O}((z-p_i)^2)(dz)^2.
        \end{align*} 
        Note that the quadratic loop equation includes the term with $\omega_{0,1}$, in contrast to the integrand of the definition of (Log-)TR \eqref{eq:TR-introLog}.
        
        Let us manipulate the right-hand side of \eqref{lemmaequationxy} by writing
        \begin{align*}
            &\frac{1}{2}\sum_{j}\mathrm{Res}_{z\to a_j}\frac{x(z)}{dx(z)}\sum_{g_1+g_2=g,\; g_i>0}\omega_{g_1,1}(z)\omega_{g_2,1}(z)\\[1mm]
            =&\frac{1}{2}\sum_{j}\mathrm{Res}_{z\to a_j}\frac{x(z)}{dx(z)}\Bigg(\sum_{g_1+g_2=g,\; g_i>0}\omega_{g_1,1}(z)\omega_{g_2,1}(z)+\omega_{g-1,2}(z,z)\Bigg),
        \end{align*}
        since $\omega_{g-1,2}$ has no pole at the log-vital singular point $a_i$ of $y$ (only the $\omega_{g,1}$ have poles at $a_i$). Next, we move the integration contour originally around all log-vital points of $y$ to encircle all the other possible poles or cuts. The possible poles are the ramification points $p_i$ of $x$ and the singularities of $x$ (including logarithmic singularities and cuts of $x$). Around a singularity of $x$, both $\omega_{g',1}$ and $\omega_{g-1,2}$ are regular, so a possible pole for the residue calculation necessarily comes from $\frac{x}{dx}\,dz$. However, around a pole where $x\sim \frac{1}{(z-b)^k}$, the term $\frac{x}{dx}\,dz$ vanishes and gives no contribution to the residue calculation. This holds in particular if the singularity is logarithmic, $k=0$. Thus, after moving the integration contour, the only singular points are the ramification points $p_i$ of $x$:
        \begin{align*}
            =-\frac{1}{2}\sum_{i}\mathrm{Res}_{z\to p_i}\frac{x(z)}{dx(z)}\Bigg(\sum_{g_1+g_2=g,\; g_i>0}\omega_{g_1,1}(z)\omega_{g_2,1}(z)+\omega_{g-1,2}(z,z)\Bigg).
        \end{align*}
        The next steps are manipulations of this expression by the linear and quadratic loop equations (exactly as in the proof of \cite[Corollary 4.1]{Eynard:2007kz}):
        \begin{align*}
            =&-\frac{1}{4}\sum_{i}\mathrm{Res}_{z\to p_i}\frac{x(z)}{dx(z)}\Bigg(\sum_{g_1+g_2=g,\; g_i>0}\omega_{g_1,1}(z)\omega_{g_2,1}(z)+\omega_{g-1,2}(z,z)\\[1mm]
            &\quad\quad\quad+\sum_{g_1+g_2=g,\; g_i>0}\omega_{g_1,1}(\sigma_i(z))\omega_{g_2,1}(\sigma_i(z))+\omega_{g-1,2}(\sigma_i(z),\sigma_i(z))\Bigg)\\[1mm]
            =&\frac{1}{2}\sum_{i}\mathrm{Res}_{z\to p_i}\frac{x(z)}{dx(z)}\Bigg(\sum_{g_1+g_2=g,\; g_i>0}\omega_{g_1,1}(z)\omega_{g_2,1}(\sigma_i(z))+\omega_{g-1,2}(z,\sigma_i(z))\Bigg)\\[1mm]
            =&-\frac{1}{2}\sum_{i}\mathrm{Res}_{z\to p_i}\frac{x(z)}{dx(z)}\Big(\omega_{0,1}(z)\omega_{g,1}(\sigma_i(z))+\omega_{0,1}(\sigma_i(z))\omega_{g,1}(z)\Big)\\[1mm]
            =&\sum_{i}\mathrm{Res}_{z\to p_i}\frac{x(z)}{dx(z)}\omega_{0,1}(z)\omega_{g,1}(z)\\[1mm]
            =&\sum_{i}\mathrm{Res}_{z\to p_i}x(z)y(z)\omega_{g,1}(z).
        \end{align*}
    \end{proof}
\end{lemma}

The corresponding relation of Lemma \ref{lem:xyomegag1} in TR was of importance in the derivation of the difference of $F_g$ and $F_g^\vee$ for meromorphic $x,y$ in \cite{Eynard:2013csa}. This is why we need it for our extension to allow logarithmic poles of $x$ and $y$. Clearly, there is an $x$-$y$ dual analog of \eqref{lemmaequationxy} of the form
\begin{align}\label{lemmaequationxydual}
        \sum_i\Res_{z\to p^\vee_i} x(z)y(z)\omega^\vee_{g,1}(z)=\frac{1}{2}\sum_{j}\mathrm{Res}_{z\to a^\vee_j}\frac{y(z)}{dy(z)}\sum_{g_1+g_2=g,\; g_i>0}\omega^\vee_{g_1,1}(z)\omega^\vee_{g_2,1}(z).
\end{align}
Now we are ready to derive the main result of the article, where we want to emphasize one more time that it holds for $x,y$ including logarithmic singularities such that $dx,dy$ is meromorphic.

\begin{theorem}\label{thm:mainthmtext}
    Let $\omega_{g,n}$ be generated by Log-TR for the spectral curve $(\Sigma,x,y,B)$ and $\omega_{g,n}^\vee$ by the spectral curve $(\Sigma,y,x,B)$ where $x$ and $y$ is swapped. For $g\geq 2$, the difference of the free energies $F_g$ and $F_g^\vee$ (defined by \eqref{freeenergy2} and \eqref{freeenergydual}) is
    \begin{align}
        &(2-2g)(F_g-F_g^\vee)\\\nonumber
        =&\sum_{a_i,a_i^\vee,b_i,b_i^\vee}\mathop{\mathrm{Res}}_{z\to a_i,a_i^\vee,b_i,b_i^\vee}[\hbar^{2g-1}]dy\sum_{m\geq 2}\frac{d^{m-1}y}{dx^{m-1}}\\\nonumber
        &\times [u^m]\frac{\exp\left(\sum_{h,n}\frac{\hbar^{2h+n-2}}{n!}\int_{y-\hbar u/2}^{y+\hbar u/2}\bigg(\omega_{h,n}^\vee-\frac{\delta_{(h,n),(0,2)}dy_1dy_2}{(y_1-y_2)^2}\bigg)-x u\right)}{u\hbar}.
    \end{align}
    The set of points $\{a_i\},\{a_i^\vee\},\{b_i\},\{b_i^\vee\}$ are the points where $\omega_{0,1}$ and/or $\omega_{0,1}^\vee$ has non-vanishing residue.
    \begin{proof}
        From the definition of $F_g$ and $F_g^\vee$, we get
        \begin{align*}
    &(2-2g)(F_g-F_g^\vee)\\
    =&\sum_{p_i}\mathop{\mathrm{Res}}_{z\to p_i}\omega_{g,1}(z)\Phi(z)-\sum_{p_i^\vee}\mathop{\mathrm{Res}}_{z\to p_i^\vee}\omega_{g,1}^\vee(z)\Phi^\vee(z)\\
    =&\sum_{p_i,p_i^\vee}\mathop{\mathrm{Res}}_{z\to p_i,p_i^\vee}(\omega_{g,1}(z)\Phi(z)-\omega_{g,1}^\vee(z)\Phi^\vee(z))\\
    =&\sum_{p_i,p_i^\vee}\mathop{\mathrm{Res}}_{z\to p_i,p_i^\vee}\big[(\omega_{g,1}(z)+\omega_{g,1}^\vee(z))\Phi(z)-x(z)y(z)\omega_{g,1}^\vee(z)\big].
        \end{align*}
        We have used the fact that $\omega_{g,1}(z)$ and $\Phi(z)$ have no pole at $p_i^\vee$, and $\omega^\vee_{g,1}(z)$ and $\Phi^\vee(z)$ no pole at $p_i$. Furthermore, around those points one has $\Phi^\vee=x y-\Phi$, where a constant term would not contribute due to the fact that all $\omega^\vee_{g,1}$ is residue-free. For the last term, we  are now applying Lemma \ref{lem:xyomegag1} or rather the $x$-$y$ dual version stated in \eqref{lemmaequationxydual}. For $\omega_{g,1}(z)+\omega_{g,1}^\vee(z)$, the $x$-$y$ duality formula \eqref{xyomg1}, where the $[u^{m=0}]$ term yields the $\omega_{g,1}^\vee$-term. Thus, for $\omega_{g,1}(z)+\omega_{g,1}^\vee(z)$ the sum over $m$ starts at $m=1$ in \eqref{xyomg1}. We conclude for the expression
\begin{align*}
    =&\sum_{p_i,p_i^\vee}\mathop{\mathrm{Res}}_{z\to p_i,p_i^\vee}\Phi(z)[\hbar^{2g-1}]\sum_{m\geq 1} \left(-d\frac{1}{dx}\right)^m[u^m](-dy)\\\nonumber
        &\times\frac{\exp\left(\sum_{h,n}\frac{\hbar^{2h+n-2}}{n!}\int_{y-\hbar u/2}^{y+\hbar u/2}\bigg(\omega_{h,n}^\vee-\frac{\delta_{(h,n),(0,2)}dy_1dy_2}{(y_1-y_2)^2}\bigg)-x u\right)}{u\hbar}\\
    &-\frac{1}{2}\sum_{a_j^\vee}\mathrm{Res}_{z\to a^\vee_j}\frac{y(z)}{dy(z)}\sum_{g_1+g_2=g,\; g_i>0}\omega_{g_1,1}^\vee(z)\omega_{g_2,1}^\vee(z).
\end{align*}
For the residue calculation around $p_i,p_i^\vee$ we have to be careful with the logarithmic singularities and cuts of $\Phi(z)$. We cannot just deform the integration contour. However, since $m\geq 1$, we can integrate by parts around each point $p_i,p_i^\vee$ and have $\frac{1}{dx(z)}d\Phi(z)=y(z)$:
\begin{align}\nonumber
    =&\sum_{p_i,p_i^\vee}\mathop{\mathrm{Res}}_{z\to p_i,p_i^\vee}y(z)[\hbar^{2g-1}]\sum_{m\geq 1} \left(-d\frac{1}{dx}\right)^{m-1}[u^m](-dy)\\\nonumber
        &\times\frac{\exp\left(\sum_{h,n}\frac{\hbar^{2h+n-2}}{n!}\int_{y-\hbar u/2}^{y+\hbar u/2}\bigg(\omega_{h,n}^\vee-\frac{\delta_{(h,n),(0,2)}dy_1dy_2}{(y_1-y_2)^2}\bigg)-x u\right)}{u\hbar}\\\label{lastlineproo}
    &-\frac{1}{2}\sum_{a_j^\vee}\mathop{\mathrm{Res}}_{z\to a^\vee_j}\frac{y(z)}{dy(z)}\sum_{g_1+g_2=g,\; g_i>0}\omega_{g_1,1}^\vee(z)\omega_{g_2,1}^\vee(z).
\end{align}
For $m=1$, taking the $[u^1]$ and $[\hbar^{2g-1}]$ coefficient extracts the term 
\begin{align*}
    -\frac{\omega_{g-1,2}(z,z)+\frac{1}{2}\sum_{\substack{g_1+g_2=g\\ g_i>0}}\omega_{g_1,1}(z)\omega_{g_2,1}(z)}{dy}.
\end{align*}
The important observation is now that the $m=1$ term cancels the residue calculation around the points $a_i^\vee$ of the line \eqref{lastlineproo}. We can first add to the line \eqref{lastlineproo} a term of the form $\omega_{g-1,2}(z,z)$ which has no pole at $a_i^\vee$. Putting this together with the $m=1$ term gives
\begin{align*}
    -\frac{1}{2}\sum_{a_j^\vee,p_i,p_i^\vee}\mathop{\mathrm{Res}}_{z\to a_j^\vee,p_i,p_i^\vee}\frac{y(z)}{dy(z)}\Big(\omega_{g-1,2}^\vee(z,z)+\sum_{g_1+g_2=g,\; g_i>0}\omega_{g_1,1}^\vee(z)\omega_{g_2,1}^\vee(z)\Big)=0.
\end{align*}
The annihilation of the term comes from the fact that we actually sum over all residues. There would be no contribution from $b_i$ and $b_i^\vee$ and especially also not from $a_i$, where $\frac{y}{y'}\sim (z-a_i)\log(z-a_i)$. The cancellation of the $[u^{1}]$-term is quite important because this gives us the possibility to integrate one more time by parts, giving now
\begin{align*}
    &(2-2g)(F_g-F_g^\vee)\\
    =&\sum_{p_i,p_i^\vee}\mathop{\mathrm{Res}}_{z\to p_i,p_i^\vee}y(z)[\hbar^{2g-1}]\sum_{m\geq 2} \left(-d\frac{1}{dx}\right)^{m-1}[u^m](-dy)\\\nonumber
        &\times\frac{\exp\left(\sum_{h,n}\frac{\hbar^{2h+n-2}}{n!}\int_{y-\hbar u/2}^{y+\hbar u/2}\Big(\omega_{h,n}^\vee-\frac{\delta_{(h,n),(0,2)}dy_1dy_2}{(y_1-y_2)^2}\Big)-x u\right)}{u\hbar}\\
        =&\sum_{p_i,p_i^\vee}\mathop{\mathrm{Res}}_{z\to p_i,p_i^\vee}\frac{dy}{dx}[\hbar^{2g-1}]\sum_{m\geq 2} \left(-d\frac{1}{dx}\right)^{m-2}[u^m](-dy)\\\nonumber
        &\times\frac{\exp\left(\sum_{h,n}\frac{\hbar^{2h+n-2}}{n!}\int_{y-\hbar u/2}^{y+\hbar u/2}\Big(\omega_{h,n}^\vee-\frac{\delta_{(h,n),(0,2)}dy_1dy_2}{(y_1-y_2)^2}\Big)-x u\right)}{u\hbar}.
\end{align*}    
Just now, we are able to move safely the contour around all the other possible poles $a_i,a_i^\vee,b_i,b_i^\vee$ since $\frac{dy}{dx}=\frac{d^2}{dx^2}\Phi$ has now no logarithmic poles or any branch cut. We can also integrate $m-2$ more times for each term to derive the desired result.
    \end{proof}
\end{theorem}

The difference between $F_g$ and $F_g^\vee$ seems to be, for general spectral curves, quite complicated. It is not yet clear to us if the residue calculation around $a_i,a_i^\vee, b_i,b_i^\vee$ can actually be computed or simplified for general spectral curves. However, in important examples where TR is known to have applications, the computation of the residue can actually be performed. It turns out to be a quite non-trivial computation but still somehow straightforward. \\

A very important class of spectral curves is that in which $y$ is unramified. This means that the dual free energy $F_g^\vee$ is by definition \eqref{freeenergydual} trivial, that is, $F_g^\vee=0$ for $g\geq 2$. However, the actual free energy $F_g$ is known to be non-trivial. These classes of curves are discussed in Examples \ref{ex:omg1yzxrational}, \ref{ex:omg1ylogxrational}, \ref{ex:omg1yzxrational} and \ref{ex:omg1ylogxlog}. Inserting these examples into Theorem \ref{thm:mainthmtext}, we obtain Corollary \ref{cor:explicitFg}. 

Now, we want to make our hands dirty and apply the Corollary to actually compute the free energy $F_g$ in a situation where $y$ is unramified. We separate the discussion into two different classes of applications in the following subsections:
\begin{itemize}
    \item[Sec. \ref{sec.exrational}:] Rational curves with $x$ rational and $y=z$. The spectral curve is equivalent to the vanishing locus of a polynomial $P(x,y)=0$ in $\mathbb{C}^2$. The 1-form is $\omega_{0,1}=ydx$.
    \item[Sec. \ref{sec.exseiberg}:] $x$ has log-vital points and $u=e^x$ is rational, while $y=z$. The spectral curve is equivalent to the vanishing locus of a polynomial $P(u,y)=0$ in $\mathbb{C}^\times\times\mathbb{C}$. The 1-form is $\omega_{0,1}=ydx=y\frac{du}{u}$. This class of curves can be interpreted as some Seiberg-Witten curve.
\end{itemize}
For both classes of examples, the free energy $F_g$ coincides with the free energy defined by TR, since Log-TR reduces to TR. However, in Sec. \ref{sec.exseiberg} the dual side indeed requires the use of Log-TR.

\subsection{Free Energy of simple rational curves}\label{sec.exrational}
Let us start with the simplest non-trivial example, the Harer–Zagier curve which was already mentioned before. It takes the following form 
\begin{align*}
    x=z+\frac{1}{z},\quad y=z,\qquad B=\frac{dz_1\,dz_2}{(z_1-z_2)^2}.
\end{align*}
The free energy $F_g$ of this curve is known from its relation to the Gaussian unitary ensemble, which has an important connection to the Euler characteristic of the moduli space of complex curves. Let us mention once more that computing $F_g$ through TR would require following the recursive procedure in $2g+n-2$. Through determinantal formulas \cite{Bergere:2009zm} and/or the quantum spectral curve \cite{JEP_2017__4__845_0}, it might be possible to derive the free energy explicitly as well. We will show that this can be done via Corollary \ref{cor:explicitFg}. The only poles are at $b_1^\vee=0$ and $b_2^\vee=\infty$, and $\omega_{0,1}=ydx$ has no other pole. Thus, the free energy reduces to the residue calculation
\begin{align*}
    F_g=&\frac{1}{2-2g}\mathop{\mathrm{Res}}_{z\to 0,\infty}[\hbar^{2g-1}]\,dz\sum_{m\geq 2}\frac{d^{m-1}z}{dx(z)^{m-1}}[u^m]\frac{\exp\left[\frac{1}{\hbar}\int_{z-\hbar u/2}^{z+\hbar u/2} x(z)\,dz-x(z) u\right]}{u\hbar}\\[1mm]
        =&\frac{1}{2-2g}\mathop{\mathrm{Res}}_{z\to 0,\infty}[\hbar^{2g-1}]\,dz\sum_{m\geq 2}\frac{d^{m-1}z}{dx(z)^{m-1}}[u^m]\bigg(\frac{1+\frac{\hbar u}{2z}}{1-\frac{\hbar u}{2z}}\bigg)^{1/\hbar}\frac{e^{-u/z}}{u\hbar}.
\end{align*}
For an expansion around $z=0$, we find 
\[
\frac{d^{m-1}z}{d\Big(\frac{1}{z}+\mathcal{O}(z^0)\Big)^{m-1}}=(-1)^{m}(m-1)!z^{m}+\mathcal{O}(z^{m+2}).
\]
The expansion of 
\[
\left(\frac{1+\frac{\hbar u}{2z}}{1-\frac{\hbar u}{2z}}\right)^{1/\hbar}
\]
around $z=0$, when expanded first in $\hbar$ and then in $u$, is not trivial. The expansion of this term, see for instance \cite[Cor.3.1.8]{Lando}, has the form
\begin{align*}
    \left(\frac{1+\frac{u \hbar}{2z}}{1-\frac{u \hbar}{2z}}\right)^{1/\hbar}
    =&1+\frac{u}{z}+\sum_{n=1}^\infty\sum_{g=0}^{n}\hbar^{2g}\frac{\left(\frac{u}{z}\right)^{n+1}}{(n+1)(n-2g)!}\,[t^{2g}]\Bigg(\frac{t/2}{\tanh(t/2)}\Bigg)^{n+1}\\[1mm]
    =&e^{u/z}+\sum_{g=0}^{\infty}\hbar^{2g}[t^{2g}]\sum_{n=1}^\infty\frac{\left(\frac{u}{z}\right)^{n+1}}{(n+1)(n-2g)!}\Bigg(\frac{t/2}{\tanh(t/2)}\Bigg)^{n+1}.
\end{align*}

Let us define 
\[
A:=\frac{t/2}{\mathrm{tanh}(t/2)}=\sum_{n=0}^\infty \frac{B_{2n}\, t^{2n}}{(2n)!}.
\]
The coefficient we need to reconstruct the \(e^{u/z}\) for every order in \(\hbar\) is
\begin{align}\label{ang}
    a_{n,g}:=\sum_{i=0}^n\frac{[t^{2g}]A^{2g+i+1}}{(2g+i+1)i!}\frac{(-1)^{n-i}}{(n-i)!},
\end{align}
since then the following holds
\[
\left(\frac{1+\frac{u \hbar}{2z}}{1-\frac{u \hbar }{2z}}\right)^{1/\hbar}e^{-u/z}=\sum_{g=0}^\infty\hbar^{2g}\left(\frac{u}{z}\right)^{2g+1}\sum_{n=0}^{g-1}\left(\frac{u}{z}\right)^{n} a_{n,g}.
\]
In summary, we have after the expansions the residue computation 
\begin{align*}
   &\frac{1}{2-2g}\mathop{\mathrm{Res}}_{z\to 0}[\hbar^{2g-1}]\,dz\sum_{m\geq 2}\Big[(-1)^{m}(m-1)!z^{m}+\mathcal{O}(z^{m+2})\Big]\\[1mm]
    &\qquad\times [u^m]\left(\frac{u}{z}\right)^{2g+1}\sum_{n=0}^{g-1}\left(\frac{u}{z}\right)^{n} \frac{a_{n,g}}{u} \\
    =&\frac{1}{2-2g}\sum_{n=0}^{g-1}(-1)^{2g+n-1}(2g+n-1)!a_{n,g}\\[1mm]
    =&\frac{1}{2-2g}\frac{B_{2g}\,(2g-1)!}{(2g)!},
\end{align*}
where in the last step we have used Lemma \ref{lem:ang} together with 
\[
A=\frac{t/2}{\mathrm{tanh}(t/2)}=\sum_{n=0}^\infty \frac{B_{2n}\, t^{2n}}{(2n)!}.
\]
For the expansion around \( z = \infty \), we find  
\[
\frac{d^{m-1} z}{dx(z)^{m-1}} = (-1)^{m-1} \frac{(m-1)!}{z^m} + \mathcal{O} \left(\frac{1}{z^{m+1}} \right),
\]
but also  
\[
\left( \frac{1 + \frac{\hbar u}{2z}}{1 - \frac{\hbar u}{2z}} \right)^{1/\hbar} \frac{e^{-u/z}}{u\hbar}
\]
expands purely into negative powers of \( z \), which in total results in a vanishing contribution from the residue at \( z = \infty \).

Thus, the free energy $F_{g\geq 2}$ computed by TR for $x=z+\frac{1}{z}$ and $y=z$ is
\[
F_g= \frac{B_{2g} }{2g(2-2g)},
\]
which coincides with the expected result. \\

Note that exactly the same computation can be applied for much more general curves with still \(y=z\) but 
\[
x=z+\sum_k\frac{c_i}{z-b_i^\vee}.
\]

\begin{corollary}\label{cor:rationalx}
    Let \(y=z\) and 
    \[
    x=z+\sum_k\frac{c_i}{z-b_i^\vee}.
    \]
    The free energy \(F_{g\geq 2}\) computed by TR has the form
    \[
    F_g=\frac{B_{2g} }{2g(2-2g)}\sum_i c_i^{2-2g}.
    \]
    \begin{proof}
        The computation follows exactly the same strategy as for the Harer–Zagier curve in the discussion before. Around each singular point \(b_i^\vee\), the only contribution from \(\frac{d^{m-1}y}{dx^{m-1}}\) is of the form 
        \[
        (-1)^m(m-1)!(z-b_i^\vee)^m.
        \]
        At each singular point \(b_i^\vee\), the residue of \(\omega_{0,1}\) is \(c_i\). The rescaling property \(\omega_{0,1}\to \lambda\omega_{0,1}\) yields a rescaling of all \(\omega_{g,n}\to \lambda^{2-2g-n}\omega_{g,n}\) including 
        \[
        F_g\to \lambda^{2-2g}F_g.
        \]
        This is performed at every singular point \(b_i^\vee\). Summing over all singular points, we obtain the result.
    \end{proof}
\end{corollary}

We want to include two more examples\footnote{We want to thank Nitin Kumar Chidambaram for suggesting those examples.} which are of interest in enumerative geometry. More precisely, the following spectral curves compute \(r\)-spin \(\psi\)-class intersection numbers of the moduli space of complex curves \(\overline{\mathcal{M}}_{g,n}\) \cite{Witten:1990hr,Kontsevich:1992ti,Bouchard:2016obz,Belliard:2021jtj}, and also those for "negative" \(r\)-spin classes. These have been properly defined from an algebraic-geometric perspective by Norbury for \(r=2\) \cite{Norbury:2017eih} and further developed in \cite{Chidambaram:2022cqc}, now known as \(\Theta\)-classes. Interestingly, from the Cohomological Field Theory perspective, both curves are quite different; see \textit{op. cit.} This difference is actually visible by looking at the free energy.

\begin{proposition}
    The free energy of the \(r\)-spin curve, parametrized for instance by 
    \begin{align}
        x=\frac{z^r}{r}-\epsilon z,\qquad y=z,
    \end{align}
    with \(r\in \mathbb{Z}_{\geq 1}\), is
        $F_{g\geq 2}=0.$
        
    The free energy of the negative \(r\)-spin curve, parametrized for instance by 
    \begin{align}
        x=\frac{z^{-r}}{r}-\epsilon z^{-1},\qquad y=z,
    \end{align}
    with \(r\in \mathbb{Z}_{\geq 1}\), is independent of $r$ and explicitly given by
    \[
        F_{g\geq 2}=\frac{B_{2g}}{2g (2-2g)}\epsilon^{2-2g},
    \]
    exhibiting singular behavior at \(\epsilon= 0\).
    \begin{proof}
        Thanks to the deformation by \(\epsilon\), both curves fit into the ordinary class of TR (i.e. they have no higher order ramification points). For the first curve, the dual free energy is \(F_g^\vee=0\). Furthermore, there is no point at which the residue of \(\omega_{0,1}\) or \(\omega_{0,1}^\vee\) does not vanish. Thus, the set of points \(\{b\}\) is empty, and the free energy \(F_g\) is therefore also zero.

        For the second curve, no dual free energy arises either. There are two points where the residue of \(\omega_{0,1}\) and \(\omega_{0,1}^\vee\) does not vanish, namely \(z\in \{0,\infty\}\) with residue \(\pm \epsilon\). We need to analyze the expansion around \(z=0\) and \(z=\infty\). At \(z=0\), we find
        \[
            \frac{d^m z}{d\Bigl(\frac{1}{r \epsilon z^r}-\frac{\epsilon }{z}\Bigr)^m} = (-1)^m(1+r(m-1))_r! \, z^{r m+1}+\mathcal{O}(z^{(r+1)m}),
        \]
        where \((1+rm)_r!:= (1+rm)(1+r(m-1))\cdots (1+r)\) is the \(r\)-factorial. It is easy to show that expanding the exponential yields, for a fixed order in \(\hbar^{2g-1}\), an expansion around \(z=0\) of the form 
        \[
            [\hbar^{2g-1}]\frac{\exp\left[\frac{1}{\hbar}\int_{z-\hbar u/2}^{z+\hbar u/2} x(z)\,dz-x(z) u\right]}{u\hbar}=\sum_{n=0}^{g-1}u^{2g+n}\sum_{k=0}^{n+1}\frac{a_{g,n,k}}{z^{2g+1+n+(r-1)k}}
        \]
        for some coefficients \(a_{g,n,k}\). After multiplying by \(\frac{d^{m-1} z}{dx^{m-1}}\) with \(m=2g+n\), the residue at \(z=0\) clearly vanishes for all \(g>1\).

        For the residue around \(z=\infty\), the calculation from the Harer–Zagier curve can be repeated one-to-one (comparing with $z=0$ in Harer-Zagier). The terms \(z^r\) do not contribute at all. Thus, we obtain for the \(z\to \infty\) residue the result 
        \[
            \frac{B_{2g}}{2g (2-2g)}\epsilon^{2-2g}.
        \]
    \end{proof}
\end{proposition}

\subsection{Free Energy of simple Seiberg-Witten-like curves}\label{sec.exseiberg}
In some examples of Seiberg-Witten curves, the base curve is a punctured sphere. A polynomial describing such a curve, let's say of the form \(P(u,y)=0\) in \(\mathbb{C}^\times \times \mathbb{C}\), has the canonical 1-form \(y\frac{du}{u}\). From the perspective of TR, we would rather consider the polynomial \(P(e^x,y)=0\) with 1-form \(\omega_{0,1}=y\,dx\). The special case of spectral curves we want to consider here are curves of the form
\begin{align}\label{speccurvexlogyz}
    x=C+\sum_i \log(z-a_i^\vee),\qquad y=z, \qquad B=\frac{dz_1dz_2}{(z_1-z_2)^2}.
\end{align}
Note that for this curve Log-TR reduces to TR, since no log-vital points exist for \(y\), and thus the free energy \(F_g\) is the free energy of TR in the original sense of its definition. However, looking at the \(x\)-\(y\) dual side, Log-TR indeed needs to be considered since \(x\) has log-vital points at \(a_i^\vee\). From Corollary \ref{cor:explicitFg}, the free energy of TR can be computed by 
\begin{align*}
    F_g=&\frac{1}{2-2g}\sum_{i}\mathop{\mathrm{Res}}_{z\to a_i^\vee}[\hbar^{2g-1}]dz\sum_{m\geq 2}\frac{d^{m-1}z}{dx(z)^{m-1}}[u^m]\frac{\exp\bigg[\Big(\frac{\cS(u\hbar \partial_z)}{\cS(\hbar \partial_z)}-1\Big)x(z) u\bigg]}{u\hbar}.
\end{align*}
This simplification appears since all \(\alpha_i^\vee=1=\Res_{z\to a_i^\vee}dx\) and \(x\) consists solely of logarithms and no rational dependence. The explicit expression for the free energy needs some computational steps, but simplifies greatly to:

\begin{proposition}\label{prop:freexlogyz}
    The free energy $F_{g\geq 2}$ computed by TR of the spectral curve \eqref{speccurvexlogyz} has the compact form
    \begin{align*}
        F_g=\frac{B_{2g}}{2g(2-2g)}\sum_{i\neq j}(a_i^\vee-a_j^\vee)^{2-2g}.
    \end{align*}
\begin{proof}
    The proof follows from direct computation around each of the poles $a_i^\vee$. We will show that 
    \begin{align}\label{eqproponepole}
        \mathop{\mathrm{Res}}_{z\to a_i^\vee}[\hbar^{2g-1}]dz\sum_{m\geq 2}\frac{d^{m-1}z}{dx(z)^{m-1}}[u^m]\frac{\exp\bigg[\big(\frac{\cS(u \hbar \partial_z)}{\cS(\hbar \partial_z)}-1\big)x(z) u\bigg]}{u\hbar}=\frac{B_{2g}}{4g}\sum_{ j\neq i}(a_i^\vee-a_j^\vee)^{2-2g}
    \end{align}
    from which the result follows by summation over all $a_i^\vee$. 

    The exponential factorizes and can be expanded in the following way
    \begin{align*}
        &\exp\bigg[\big(\frac{\cS(u \hbar \partial_z)}{\cS(\hbar \partial_z)}-1\big)x(z) u\bigg]
        = \prod_{j}\exp\bigg[\big(\frac{\cS(u \hbar \partial_z)}{\cS(\hbar \partial_z)}-1\big) u\log(z-a_j^\vee) \bigg]\\
        =& \prod_{j}\bigg(\frac{\hbar}{z-a_j^\vee}\bigg)^u\frac{\Gamma(\frac{1+u}{2}+\frac{z-a_j^\vee}{\hbar})}{\Gamma(\frac{1-u}{2}+\frac{z-a_j^\vee}{\hbar})}\\
        =& \prod_{j}\sum_{n_j\geq 0}u(u-1)...(u-n_j+1)\left(\frac{\hbar}{z-a_j^\vee}\right)^{n_j}[t_j^{n_j}]\frac{1}{\cS(t_j)^{u+1}},
    \end{align*}
    where this rewriting was for instance used in \cite{Hock:2022pbw} based on the well-known asymptotic expansion of the $\Gamma$-function at infinity in terms of Bernoulli numbers appearing in $\cS(t)=\frac{e^{t/2}-e^{-t/2}}{t}$.

    The expansion of the exponential is inserted, one can integrate by parts for each $m$ and literally replace $u$ by $\partial_x$. The only contribution for a possible residue calculation can occur if $n_i>0$. It turns out that $n_i>0$ and at most \textit{one other} $n_k>0$ with $k\neq i$ considering that $m\geq 2$. Otherwise the action of $(-\partial_x-1)...(-\partial_x-n_i+1)\frac{f(z)}{(z-a_i^\vee)^{n_i-1}}$ is regular at $z=a_i^\vee$ by iteratively applying Lemma \ref{lem:degreeshift}.

    Also the expansion of $[t_k^{n_k}]\frac{1}{\cS(t_k)^{u+1}}$ in $u$ for higher order in $u$ vanishes after applications of Lemma \ref{lem:degreeshift}. Thus,
    we are left with
    \begin{align*}
        &\mathop{\mathrm{Res}}_{z\to a_i^\vee}[\hbar^{2g-1}]dz\sum_{m\geq 2}\frac{d^{m-1}z}{dx(z)^{m-1}}[u^m]\sum_{k\neq i}\sum_{n_i+n_k=2g}(u-1)...(u-n_i+1)\\
        \times &u(u-1)...(u-n_k+1)\frac{[t_i^{n_i}]\frac{1}{\cS(t_i)^{1}}}{(z-a_i^\vee)^{n_i}}\frac{[t_k^{n_k}]\frac{1}{\cS(t_k)^{1}}}{(z-a_k^\vee)^{n_k}}.
    \end{align*}
   Again by Lemma \ref{lem:degreeshift}, we find only contribution if \((u-1)...(u-n_k+1)\to (-1)^{n_k-1}(n_k-1)!\). We add and subtract the case of \(m=1\), shift the summation index \(m\) and find finally with Lemma \ref{lem:residuecompute} (where \(g(z)=z\) and \(f(z)=\frac{1}{(z-a_k^\vee)^{n_k}}\))
    \begin{align*}
        &\sum_{\substack{n_i+n_k=2g\\ n_k>0}}\sum_{k\neq i}(-1)^{n_k-1}(n_k-1)![t_i^{n_i}]\frac{1}{\cS(t_i)}[t_k^{n_k}]\frac{1}{\cS(t_k)}\mathop{\mathrm{Res}}_{z\to a_i^\vee}[\hbar^{2g-1}]dz\\&\times\sum_{m\geq 0}\frac{d^{m}z}{dx(z)^{m}}[u^m](u-1)...(u-n_i+1)
        \frac{1}{(z-a_i^\vee)^{n_i}}\frac{1}{(z-a_k^\vee)^{n_k}}\\
        &-\sum_{\substack{n_i+n_k=2g\\ n_k>0}}\sum_{k\neq i}[t_i^{n_i}]\frac{(-1)^{n_i-1}}{\cS(t_i)}[t_k^{n_k}]\frac{(-1)^{n_k-1}(n_k-1)!}{\cS(t_k)}\frac{d^{n_i-1}}{dz^{n_i-1}}\frac{z}{(z-a_k^\vee)^{n_k}}\Bigg\vert_{z=a_i^\vee}\\
        =&-\sum_{\substack{n_i+n_k=2g\\ n_k>0}}\sum_{k\neq i}(2g-2)![t_i^{n_i}]\frac{1}{\cS(t_i)}[t_k^{n_k}]\frac{1}{\cS(t_k)}\frac{a_i^\vee}{(a_i^\vee-a_k^\vee)^{2g-1}}\\
        &+\sum_{\substack{n_i+n_k=2g\\ n_k>0}}\sum_{k\neq i}(2g-2)![t_i^{n_i}]\frac{1}{\cS(t_i)}[t_k^{n_k}]\frac{1}{\cS(t_k)}\frac{a_i^\vee}{(a_i^\vee-a_k^\vee)^{2g-1}}\\
        &+\sum_{\substack{n_i+n_k=2g\\ n_k>0}}\sum_{k\neq i}(2g-3)![t_i^{n_i}]\frac{1}{\cS(t_i)}[t_k^{n_k}]\frac{1}{\cS(t_k)}\frac{n_i-1}{(a_i^\vee-a_k^\vee)^{2g-2}}\\
        =&(2g-3)!(g-1)[t^{2g}]\frac{1}{\cS(t)^2}\sum_{k\neq i}\frac{1}{(a_i^\vee-a_k^\vee)^{2g-2}}.
    \end{align*}
    In the last step, the summation over \(n_k+n_i\) was symmetrized. Inserting \(\frac{1}{\cS(t)^2}=\sum_n\frac{B_{2g}}{(2g-2)!2g}\), we proved \eqref{eqproponepole} and therefore the final result.
\end{proof}
\end{proposition}

The resulting free energy of Proposition \ref{prop:freexlogyz} can also be interpreted as the free energy of a completely different curve by the following consideration. Let us look at the curve defined in terms of \((\tilde{x},\tilde{y})\) defined from \((x,y)\) of \eqref{speccurvexlogyz} by the following relation:
\begin{align}\label{spectilde}
    \tilde{y}=\frac{y}{e^x},\qquad \tilde{x}=e^x
\end{align}
or equivalently,
\begin{align*}
    \tilde{x}\tilde{y}=y,\qquad x=\log \tilde{x}.
\end{align*}
Note first of all that the formal symplectic form \(dx\wedge dy=d\tilde{x}\wedge d\tilde{y}\) is invariant. Next, the 1-form is also invariant
\begin{align*}
    \omega_{0,1}=y\,dx=\tilde{y}\tilde{x}\,d(\log \tilde{x})=\tilde{y}\,d\tilde{x}.
\end{align*}
The Bergman kernel is also the same, \(B=\omega_{0,2}=\frac{dz_1dz_2}{(z_1-z_2)^2}\). Finally, all ramification points \(p_i\) of \(x\) coincide with the ramification points of \(\tilde{x}\) since 
\[
dx=d\log\tilde{x}=\frac{d\tilde{x}}{\tilde{x}}.
\]
Therefore, all \(\omega_{g,n}\) generated by TR of the curve \eqref{speccurvexlogyz} coincide with \(\omega_{g,n}\) generated by TR of the curve \eqref{spectilde}. This is passed on to the free energy \(F_{g\geq2}\). 

As mentioned before, this type of curves appears, for instance, in Seiberg-Witten theory or in Hurwitz theory, and both representations in terms of $(x,y)$ or $(\tilde{x},\tilde{y})$ are used in the literature. Considering this equivalence, we can prove a conjecture by Borot et al. \cite{Borot:2024uos} about the free energy of their so-called Gaiotto curve. For our understanding the name Gaiotto curve seems a bit misleading, since the Gaiotto curve is usually called the base curve in the physics community. The following curve is rather a degenerate Seiberg-Witten curve over the Gaiotto curve. However, this confusing naming has nothing to do with the actual conjecture of \cite{Borot:2024uos}. Applying Proposition \ref{spectilde}, we conclude:

\begin{corollary}[Conjecture in \cite{Borot:2024uos}]
    The spectral curve 
    \begin{align*}
        \prod_{i=1}^r\bigg(\tilde{y}-\frac{Q_a}{\tilde{x}}\bigg)+\frac{(-\Lambda)^r}{\tilde{x}^{r+1}}=0
    \end{align*}
    computes via TR the free energy 
    \begin{align*}
        F_{g\geq 2}=-\frac{B_{2g}}{2g(2g-2)}\sum_{i\neq j}(Q_i-Q_j)^{2-2g}
    \end{align*}
    which coincides with the perturbative part (in the \(\Omega\)-background) of the Nekrasov partition function of \(\mathcal{N}=2\) pure supersymmetric gauge theory \cite{Nekrasov:2003rj}.
    \begin{proof}
        This curve has a parametrization of the form 
        \begin{align*}
            \tilde{x}=-\frac{\Lambda^r}{\prod_{a=1}^r (Q_a-z)},\qquad \tilde{y}=-\frac{z}{\Lambda^r}\prod_{a=1}^r(Q_a-z).
        \end{align*}
        Defining \(y=\tilde{y}\tilde{x}=z\) and \(x=\log\tilde{x}=r\log \Lambda +\sum_a \log (z-Q_a)\) brings this curve in the form of Proposition \ref{prop:freexlogyz}. Both curves have the same free energy for \(g\geq 2\) as explained before. The log-vital points of \(x\) are identified by \(a_i^\vee=Q_i\).
    \end{proof}
\end{corollary}
Another curve of interest in Hurwitz theory related to Whittaker vector, \(\mathcal{W}\)-algebras and (shifted) Airy structure can also be considered \cite{Borot:2021btb,chidambaram2024bhurwitznumberswhittakervectors}.
\begin{corollary}
    The (so-called CDO) spectral curve 
    \[
        \prod_{a=1}^r\Bigg(\frac{P_a}{\tilde{x}}+\tilde{y}\Bigg)+\frac{1}{\Lambda^r}\prod_{a=1}^{r-1}\Bigg(\frac{Q_a}{\tilde{x}}-\tilde{y}\Bigg)=0
    \]
    computes via TR the free energy 
    \[
        F_{g\geq 2}=\frac{B_{2g}}{2g(2-2g)}\left(\sum_{i\neq j}(Q_i-Q_j)^{2-2g}+\sum_{i\neq j}(P_i-P_j)^{2-2g}-\sum_{i,j}(Q_i+P_j)^{2-2g}\right).
    \]
    \begin{proof}
        This curve has a parametrization of the form 
        \[
            \tilde{x}=-\Lambda^r\frac{\prod_{a=1}^r (P_a+z)}{\prod_{a=1}^{r-1} (Q_a-z)},\qquad \tilde{y}=\frac{z}{\tilde{x}}.
        \]
        Defining 
        \[
            y=\tilde{y}\tilde{x}=z\quad\text{and}\quad x=\log\tilde{x},
        \]
        we see that the multidifferentials, the ramification points and hence the free energy are preserved. In this way the curve is brought into the form of Proposition~\ref{prop:freexlogyz}, which immediately implies the stated result.
    \end{proof}
\end{corollary}

The conjecture of \cite{Borot:2024uos}, which is proved here, provides an important step towards understanding the connection between TR and Seiberg-Witten theory as well as \(\mathcal{N}=2\) supersymmetric gauge theory. \\

Interestingly, the free energy obtained in Corollary~\ref{cor:rationalx} in the previous subsection has exactly the same form as the free energy of Proposition~\ref{prop:freexlogyz} after identifying \(a_i^\vee-a_j^\vee\) with the parameters \(c_k\). It is not clear to us whether this is just a coincidence or if both curves are actually related by some symplectic transformation. Similar representations of the free energy were also derived for very specific curves in \cite{10.1093/integr/xyz004}.

\section{Concluding comments}
In this article, we have taken one step forward into understanding the $x$-$y$ duality and (non-)symplectic invariance of the free energy in the theory of TR. In doing so, we have used the extended version Log-TR which follows the general duality transformation known as $x$-$y$ duality for the multidifferentials $\omega_{g,n}$. We have applied the general difference formula between $F_g$ and $F_g^\vee$ to actually compute $F_g$ in case of $F_g^\vee=0$. This has proved an open conjecture by Borot et al. \cite{Borot:2024uos} about their so-called Gaiotto spectral curve which should coincide with the perturbative expansion of the Nekrasov partition function in the $\Omega$-background. 

A class of examples not discussed in very details here are the mirror curves. Equation \eqref{coreqylog} of Corollary \ref{cor:explicitFg} can be applied for instance to the topological vertex curve \cite{Aganagic:2003db} or the resolved conifold. This was done by the author with a computer algebra system and we were able to confirm all examples we tried that we derived by \eqref{coreqylog} the asymptotic of the MacMahon function confirming \cite{Bouchard:2011ya,Bouchard:2011fm,Zhu:2011hs}. As explicit computations of Sec. \ref{sec.exrational} and \ref{sec.exseiberg} have shown, the actual derivation of the free energy by Corollary \ref{cor:explicitFg} can be quite tricky. We want to postpone computations of the resolved conifold to the future, having in mind the connection to torus knots \cite{Aganagic:2012jb}. It might be possible to prove that the closed string amplitude for any torus knot coincide with the closed string amplitude of the resolved confiold up to a symmetry factor by considering the symplectic transformation used in \cite{Brini:2011wi}.

A further striking research direction concerns the resummation of the free energy in the \(\hbar\)-expansion. The free energy grows factorially such that \(\hbar\) has a vanishing radius of convergence. However, all instanton contributions can be computed in some examples, for instance in \(\mathcal{N}=2\) supersymmetric gauge theory \cite{Nekrasov:2003rj}. It should be possible to derive those contributions from TR. It is quite striking that in the proof of Proposition \ref{prop:freexlogyz}, the appearing \(\Gamma\)-function on the \(x\)-\(y\) dual side takes exactly the form of the instanton contributions in \(\mathcal{N}=2\) supersymmetric gauge theory where the \(Q_i\) are shifted by \(\hbar k\) with \(k\in \mathbb{Z}\). \(Q_i\) also has the interpretation as a period. For the resummation, there is a discrete Fourier transform shifting the periods, see for instance \cite{Iwaki:2019zeq}. Thus, the \(x\)-\(y\) duality provides a new technique to tackle the non-perturbative regime in the theory of TR and to answer important questions, and even construct non-perturbative contributions. One example of tackling the non-perturbative regimes was recently realized in \cite{Alexandrov:2024qfe}, where the authors have proved with the help of the \(x\)-\(y\) duality that non-perturbative TR gives rise to a KP-\(\tau\) function, settling a longstanding question in TR.

We want to finish with a new suggestion about the redefinition of the free energy to an actual \(x\)-\(y\) invariant version. However, this might break the symplectic invariance under certain other symplectic transformations. Our suggestion is to separate the contributions of the difference of \(F_g\) in Theorem \ref{thm:mainthmtext} depending on whether the residue is taken around \(a_i\) and \(b_i\) or \(a_i^\vee\) and \(b_i^\vee\). If the residue calculation around the points \(a_i,b_i\) is subtracted from the free energy \(F_g\), and that around \(a_i^\vee,b_i^\vee\) from \(F_g^\vee\), we define a free energy that is invariant under \(x\)-\(y\) swap
\begin{align*}
    F_g-&\frac{1}{2-2g}\sum_{a_i,b_i}\mathop{\mathrm{Res}}_{z\to a_i,b_i}[\hbar^{2g-1}]dy\sum_{m\geq 2}\frac{d^{m-1}y}{dx^{m-1}}\\\nonumber
        &\times [u^m]\frac{\exp\left(\sum_{h,n}\frac{\hbar^{2h+n-2}}{n!}\int_{y-\hbar u/2}^{y+\hbar u/2}\Big(\omega_{h,n}^\vee-\frac{\delta_{(h,n),(0,2)}dy_1dy_2}{(y_1-y_2)^2}\Big)-x u\right)}{u\hbar}.
\end{align*}
This definition has the advantage that the free energy of the Harer–Zagier curve is still given by \(\frac{B_{2g}}{2g(2-2g)}\), coinciding with the dual free energy, rather than trivializing it as in \cite{Eynard:2013csa}. Furthermore, numerical computations have shown that also the correct free energy is calculated for the topological vertex even if a non-generic framing was taken. This resolves the issue observed in \cite{Bouchard:2011ya}. However, extending the free energy by the residue terms provides a definition which is not invariant under all symplectic transformations. To verify this, let us take the trivial curve \(x=z=y\) which has clearly trivial free energy \(F_g=F_g^\vee=0\). The transformation \(x\to x\) and \(y\to y+\frac{1}{x}\) transforms to a curve with trivial free energy \(F_g\) but nontrivial \(F_g^\vee\) since this would correspond to the \(x\)-\(y\) dual Harer–Zagier curve which has \(F_g=\frac{B_{2g}}{2g(2-2g)}\). Thus, a symplectic transformation generating a singular point as in this example at \(x=z=0\) may yield a nontrivial impact on the free energy, which is actually quite similar to the impact of logarithmic singularities on the \(x\)-\(y\) duality for which the extension to Log-TR was necessary.

\appendix
\section{Collection of identities}
We list some computational identities which were used for the derivation of free energies.
\begin{lemma}\label{lem:ang}
    For a formal power series \(A=1+\sum_{k\geq 1} b_k t^{2k}\), define 
    \[
        a_{n,g}=\sum_{i=0}^n\frac{[t^{2g}]A^{2g+i+1}}{(2g+i+1)}\frac{(-1)^{n-i}}{i!(n-i)!}.
    \]
    Then, the following holds
    \[
        \sum_{n=0}^{g-1}(-1)^{2g+n-1} (2g+n-1)!a_{n,g}=-b_g\, (2g-1)!.
    \]
    \begin{proof}
        We prove it by direct computation. Interchanging the sum over \(i\) in \(a_{n,g}\) with the sum over \(n\). Then we use some well-known identities for binomial coefficients to get
    \begin{align*}
        &\sum_{n=0}^{g-1}(-1)^{2g+n-1} \frac{(2g+n-1)!}{(2g-1)!}\sum_{i=0}^n\frac{[t^{2g}]A^{2g+i+1}}{(2g+i+1)i!}\frac{(-1)^{n-i}}{(n-i)!}\\[1mm]
        =&-\sum_{i=0}^{g-1}\frac{[t^{2g}]A^{2g+i+1}}{(2g+i+1)i!}\frac{(-1)^{i}}{(2g-1)!}\sum_{n=0}^{g-1}\frac{(2g+n-1)!}{(n-i)!}\\[1mm]
        =&-\sum_{i=0}^{g-1}\frac{[t^{2g}]A^{2g+i+1}}{(2g+i+1)i!}\frac{(-1)^{i}}{(2g-1)!}\frac{(3g-1)!}{(2g+i)(g-i-1)!}\\[1mm]
        =&-\frac{(3g-1)!}{(2g-1)!}[t^{2g}]\Bigg(A\int_0^A \frac{dA}{A}-\int_0^A dA\Bigg)\sum_{i=0}^{g-1}\frac{A^{2g+i} (-1)^i}{i!(g-1-i)!}\\[1mm]
        =&-\frac{(3g-1)!}{(2g-1)!(g-1)!}[t^{2g}]\Bigg(A\int_0^A \frac{dA}{A}-\int_0^A dA\Bigg)A^{2g}(A-1)^{g-1}\\[1mm]
        =&-\frac{(3g-1)!}{(2g-1)!(g-1)!}[t^{2g}]\Bigg(A\frac{(2g-1)!(g-1)!}{(3g-1)!}+\mathcal{O}(t^{2g+2})\Bigg)\\[1mm]
        =&-b_g.
    \end{align*}
    The integral is estimated by the Euler beta function due to the fact that \(A=1+b_2 t^2+\cdots\) and \((A-1)^{g-1}\in \mathcal{O}(t^{g-2})\).
    \end{proof}
\end{lemma}

\begin{lemma}\label{lem:degreeshift}
    Let \(x(y)=\text{const.}-\sum_b \log(Q_b-y)\) and 
    let \(f(y)\) be holomorphic at \(y=Q_a\). Then the differential operator reduces the order of the leading pole:
    \[
        \Bigg(\frac{d}{dx}-(n_a-1)\Bigg)\frac{f(y)}{(Q_a-y)^{n_a-1}}=\mathcal{O}\big((y-Q_a)^{-n_a+2}\big).
    \]
    \begin{proof}
        Direct calculation shows
        \begin{align*}
            &\Bigg(\frac{\prod_q(Q_q-y)}{\sum_b\prod_{q\neq b}(Q_q-y)}\frac{d}{dy}\frac{f(y)}{(Q_a-y)^{n_a-1}}-\frac{f(y)(n_a-1)}{(Q_a-y)^{n_a-1}}\Bigg)\\[1mm]
            =\,&(n_a-1)\frac{\prod_{q\neq a}(Q_q-y)}{\sum_b\prod_{q\neq b}(Q_q-y)}\frac{f(y)}{(Q_a-y)^{n_a-1}}-\frac{\prod_{q\neq a}(Q_q-y)}{\sum_b\prod_{q\neq b}(Q_q-y)}\frac{f'(y)}{(Q_a-y)^{n_a-2}}\\[1mm]
            &\quad-\frac{f(y)(n_a-1)}{(Q_a-y)^{n_a-1}}\\[1mm]
            =\,&(n_a-1)\frac{f(y)}{(Q_a-y)^{n_a-1}}\Bigg(\frac{\prod_{q\neq a}(Q_q-y)}{\sum_b\prod_{q\neq b}(Q_q-y)}-1\Bigg)
            -\frac{\prod_{q\neq a}(Q_q-y)}{\sum_b\prod_{q\neq b}(Q_q-y)}\frac{f'(y)}{(Q_a-y)^{n_a-2}}.
        \end{align*}
        The expression in the parentheses vanishes linearly as \(y\to Q_a\) since it can be simplified to 
        \[
            -\frac{(Q_a-y)\sum_{b\neq a}\prod_{q\neq a,b}(Q_q-y)}{\sum_b\prod_{q\neq b}(Q_q-y)}.
        \]
    \end{proof}
\end{lemma}

\begin{lemma}\label{lem:residuecompute}
    Let \(x(y)=\text{const.}-\sum_b \log(Q_b-y)\) and 
    let \(f(y),g(y)\) be holomorphic at \(y=Q_a\). Then we have the explicit residue computation
    \[
        \mathop{\mathrm{Res}}_{y\to Q_a}\frac{f(y)}{(y-Q_a)^{n_a}}\Big((\partial_x+(n_a-1))(\partial_x+(n_a-2))\cdots(\partial_x+1)g(y)\Big)
        =g(Q_a)f^{(n_a-1)}(Q_a),
    \]
    where \(f^{(n_a-1)}\) denotes the \((n_a-1)\)th derivative of \(f\).
    \begin{proof}
        The expansion of 
        \[
        (\partial_x+(n_a-1))(\partial_x+(n_a-2))\cdots(\partial_x+1)g(y)
        \]
        around \(Q_a\) reads
        \[
            g(Q_a)+\mathcal{O}\big((y-Q_a)^{n_a}\big),
        \]
        which can easily be proven by induction on \(n_a\). Inserting this expansion, the lemma follows immediately by performing the remaining residue calculation.
    \end{proof}
\end{lemma}

\input{main.bbl}

\end{document}

%% file: main.bbl
\newcommand{\etalchar}[1]{$^{#1}$}